\documentclass[greek,english]{aa}  

\usepackage{graphicx}
\graphicspath{{figs/}}
\usepackage{booktabs,tabularx}
\usepackage{float}
\usepackage{txfonts}
\usepackage{nicefrac,xfrac}
\usepackage{upgreek}
\usepackage{mathrsfs}

\usepackage[breaklinks=true]{hyperref}
\usepackage{xcolor}
\hypersetup{
	colorlinks   = true, 
	urlcolor     = blue, 
	linkcolor    = blue, 
	citecolor   = blue 
}
\usepackage[all]{hypcap} 

\makeatletter
\newcommand\footnoteref[1]{\protected@xdef\@thefnmark{\ref{#1}}\@footnotemark}
\makeatother
\newcolumntype{Y}{>{\centering\arraybackslash}X}

\begin{document}

	\title{A 3.5 Mpc-long radio relic in the galaxy cluster ClG~0217+70}
	
	\author{D. N. Hoang          \inst{\ref{Hamburg}} 
		\and
		X. Zhang                 \inst{\ref{Leiden},\ref{SRON}}
		\and
		C. Stuardi        \inst{\ref{Bologna},\ref{INAF}}
		\and
		T. W. Shimwell \inst{\ref{ASTRON},\ref{Leiden}}          
		\and
		A. Bonafede      \inst{\ref{Bologna},\ref{INAF}} %\fnmsep\thanks{Just to show the usage of the elements in the author field}
		\and
		M. Br\"uggen          \inst{\ref{Hamburg}} 
		\and
		G. Brunetti \inst{\ref{INAF}}
		\and
		A. Botteon \inst{\ref{Leiden}}
		\and
		R. Cassano \inst{\ref{INAF}}
		\and
		F. de Gasperin \inst{\ref{Hamburg},\ref{INAF}}
		\and
		G. Di Gennaro \inst{\ref{Hamburg},\ref{Leiden}}
		\and
		M. Hoeft \inst{\ref{TLS}}
		\and
		H. Intema \inst{\ref{Leiden}}
		\and
		K. Rajpurohit \inst{\ref{Bologna}}
		\and
		H. J. A. R\"ottgering \inst{\ref{Leiden}}  
		\and
		A. Simionescu \inst{\ref{SRON},\ref{Leiden},\ref{Kavli}}
		\and
		R. J. van Weeren \inst{\ref{Leiden}}
	}
	
	\institute{Hamburger Sternwarte, University of Hamburg, Gojenbergsweg 112, 21029 Hamburg, Germany\label{Hamburg}\\
		\email{hoang@hs.uni-hamburg.de}
		\and
		Leiden Observatory, Leiden University, PO Box 9513, NL-2300 RA Leiden, The Netherlands \label{Leiden}
		\and
		SRON Netherlands Institute for Space Research, Sorbonnelaan 2, 3584 CA Utrecht, The Netherlands \label{SRON}
		\and
		Dipartimento di Fisica e Astronomia, Universit\"a di Bologna, via Gobetti 93/2, 40122 Bologna, Italy \label{Bologna}
		\and
		INAF - Istituto di Radioastronomia di Bologna, Via Gobetti 101, I-40129 Bologna, Italy \label{INAF}
		\and
		Netherlands Institute for Radio Astronomy (ASTRON), P.O. Box 2, 7990 AA Dwingeloo, The Netherlands \label{ASTRON}
		\and
		Kavli Institute for the Physics and Mathematics of the Universe (WPI), The University of Tokyo, Kashiwa, Chiba 277-8583, Japan \label{Kavli}
		\and 
		Harvard-Smithsonian Center for Astrophysics, 60 Garden Street, Cambridge, MA 02138, USA \label{Harvard}
		\and
		Th\"uringer Landessternwarte, Sternwarte 5, 07778 Tautenburg, Germany \label{TLS}
	}
	
	\date{Received: May 30, 2021; accepted: October 02, 2021}
	
	\abstract
	{Mega-parsec scale radio sources in the form of halos and relics are often detected in dynamically disturbed galaxy clusters. Although they are associated with merger-induced turbulence and shocks, respectively, their formation is not fully understood.}
	{We aim to identify the mechanisms responsible for particle acceleration and magnetic field amplification in the halo and relics of the galaxy cluster ClG~0217+70.}
	{We observed ClG~0217+70 with LOFAR at 141~MHz and VLA at 1.5~GHz, and combine these observations with VLA 1.4~GHz archival data to study the morphological and spectral properties of the diffuse sources. We add \emph{Chandra} archival data to examine the thermal and non-thermal properties of the halo.}
	{Our LOFAR and VLA data confirm the presence of a giant radio halo in the cluster centre and multiple relics in the outskirts. The radio and X-ray emission from the halo are correlated, implying a tight relation between the thermal and non-thermal components. The diffuse radio structure in the south east with a projected size of 3.5~Mpc is the most extended radio relic detected to date. The spectral index across the relic width steepens towards the cluster centre, suggesting the electron ageing in the post-shock regions. The shock Mach numbers for the relics derived from the spectral index map range between 2.0 and 3.2. However, the integrated spectral indices lead to increasingly high Mach numbers for the relics farther from the cluster centre. This discrepancy could be because the relation between injection and integrated spectra does not hold for distant shocks, suggesting that the cooling time for the radio-emitting electrons is longer than the crossing time of the shocks. The variations in the surface brightness of the relics and the low Mach numbers imply that the radio-emitting electrons are re-accelerated from fossil plasma that could originate in active galactic nuclei.}
	{}
	
	\maketitle
	%
	%-------------------------------------------------------------------
	
	%%%%%%%%%%%%%%%%%%%%%%%%%%%%%%%%%%%%%%%
	\section{Introduction}
	\label{sec:intro}
	
	Galaxy clusters reside at the intersection of cosmic filaments and grow through a series of mergers with smaller clusters/groups of galaxies. The merging of galaxy clusters (within a few Gyrs) converts an enormous amount of gravitational energy (i.e. up to $10^{64}$~ergs) into thermal and non-thermal energy. The thermal energy is produced by heating and compression of the X-ray emitting intra-cluster medium (ICM). The non-thermal energy is channelled through the acceleration of cosmic rays (CRs) and the amplification of large-scale magnetic fields. These events are the most energetic events in the Universe after the Big Bang \citep{Sarazin2002}. During the mergers, shock waves and turbulence are generated in the ICM and are traced by diffuse radio synchrotron sources, namely radio halos and relics, respectively (see \citealt{VanWeeren2019a} and \citealt{Brunetti2014} for recent reviews). Radio halos are steep-spectrum\footnote{$S\propto\nu^\alpha$, where $S$ is flux density at frequency $\nu$ and $\alpha$ is spectral index.} ($\alpha\lesssim-1$), unpolarised, Mpc-scale sources that are detected at the cluster centre, whereas radio relics are steep-spectrum, highly-polarised at GHz frequencies, elongated sources that are normally detected in the cluster outskirts. 
	
	The favoured model for the formation of halos is the turbulent \mbox{re-acceleration} model in which fossil relativistic electrons ($\gamma\sim10^2$ prior to the \mbox{re-acceleration}) in the cluster magnetic fields are \mbox{re-accelerated} by turbulence generated during mergers and, subsequently, emit synchrotron radiation \citep{Brunetti2001,Petrosian2001a,Brunetti2007a,Brunetti2011,Brunetti2016,Pinzke2017a}. The relativistic electrons in relics were thought to be accelerated directly from the thermal pool at collisionless plasma shock fronts through Fermi-I Diffusive Shock Acceleration \citep[DSA;][]{Bell1978,Drury1983,Blandford1987,Ensslin1998,Roettiger1999a}. Despite its success in explaining the observed properties (i.e. morphological, spectral and polarisation properties) of radio relics, the DSA model requires an acceleration efficiency that is unrealistically high to explain the observed surface brightness in some relics that are generated with low-Mach number shocks (i.e. $\mathscr{M}\approx1.5-3$) \citep[e.g.][]{Hoeft2007a,Vazza2014,Botteon2020}. An alternative Diffusive Shock \mbox{Re-Acceleration} (DSRA) model where a population of relativistic fossil plasma is required to be present prior to the \mbox{re-acceleration} by the passage of the low-Mach number shocks has been proposed to solve this energy problem  \citep[][]{Markevitch2005,Kang2011a,Kang2012}. There has been observational evidence for this re-acceleration scenario in some clusters \citep[e.g.][]{weeren2013,Bonafede2014,shimwell2015,Botteon2016a,VanWeeren2017, Hoang2018, Gennaro2018, Stuardi2019}.
	
	The galaxy cluster ClG0217+70 (hereafter ClG0217; $z=0.18$, \citealt{Zhang2020}) was studied with the Very Large Array (VLA) at 1.4~GHz and 325~MHz by \cite{Brown2011a} after the first detection with the Westerbork Northern Sky Survey \citep[WENSS;][]{Rengelink1997,Delain2006,Rudnick2006}. \cite{Brown2011a} found that the cluster hosts multiple radio diffuse sources including a radio halo in the cluster centre and multiple relics on opposite sides of the cluster and internal filaments. The presence of these diffuse radio sources is consistent with the picture that the cluster is dynamically disturbed as it is suggested by the elongating morphology of the X-ray emission in the ROSAT All-Sky Survey (RASS) image \citep{Brown2011a}.
	
	\cite{Brown2011a} also analysed the photometric redshifts from galaxies in the field using the Sloan Digital Sky Survey (SDSS) optical data and found that the cluster has a redshift of $z=0.0655$, but also suggested the need of deeper optical data to verify the result. With this low-redshift estimate, \cite{Brown2011a} pointed out that the power of the radio halo in ClG0217 is an order of magnitude higher than the predicted value from the radio power--X-ray luminosity correlation \citep[][]{Liang2000,Cassano2006, Brunetti2009}. However, recent work by \cite{Zhang2020} estimated a significantly higher redshift for the cluster, i.e. $z=0.18$, from the Fe~xxv~He$\alpha$ X-ray emission line that they detected in \textit{Chandra} archival data. Using this new redshift, they found that the 1.4~GHz power of the radio halo and the $0.1-2.4$~keV X-ray luminosity within $R_{500}$\footnote{$R_n$ is the radius in which the integrated mean density of the cluster is $n$ times the critical density of the Universe at the cluster redshift.} are consistent with the radio power--X-ray luminosity relation.
	
	The mass of ClG0217 is $M_{500}=(1.06 \pm 0.11)\times10^{15}\,{\rm M_\odot}$, estimated using the $M_{500}-kT$ scaling relation \citep{Arnaud2007}, making it one of the most massive clusters known. At the locations of the northern and southern radio halo edges (see Figure~1 of \citealt{Zhang2020}), X-ray surface brightness (SB) discontinuities with density jumps of  $C_{\rm N}=1.40\pm0.16$ and $C_{\rm S}=3.0\pm0.6$, respectively were found \citep{Zhang2020}. The radio emission at 1.4~GHz is found to rapidly steepen at the location of the northern X-ray SB jump towards the north. Due to the low integration time of the X-ray observations, it is not clear whether the discontinuities are caused by shocks or cold fronts. 
	
	In this paper, we present LOw Frequency ARray \citep[LOFAR;][]{VanHaarlem2013} 120--168~MHz and VLA 1--2~GHz (C-array) observations of ClG0217 to study the diffuse radio emission from the cluster. We combined the data with the VLA 1.37--1.48 GHz (D-array) archival data to examine the spectral properties of these sources. We did not include the relatively shallow VLA \textit{P}-band data reported in \cite{Brown2011a} to this study. In addition, \textit{Chandra} X-ray archival data was used to examine the correlation between X-ray thermal and radio non-thermal emission from the radio halo. Throughout the paper, we assume a Lambda Cold Dark Matter ($\Lambda$-CDM) cosmology with $\Omega_M=0.3$, $\Omega_\Lambda=0.7$, and $H_0=70$~km~s$^{-1}$~Mpc$^{-1}$. At the cluster redshift ($z=0.18$), an angular scale of $1\arcmin$ then corresponds to a physical size of $182$~kpc.
	
	%%%%%%%%%%%%%%%%%%%%%%%%%%%%%%%%%%%%%%%
	\section{Observations and data reduction}
	\label{sec:data}
	
	\subsection{LOFAR data}
	\label{sec:lofar}
	
	\begin{table*}
		\centering
		\caption{Observation details}
		\begin{tabularx}{\textwidth}{@{}XYYY@{}}
			\hline\hline
			Telescope                  &    LOFAR 141~MHz     &          VLA 1.4~GHz          &   VLA 1.5~GHz   \\ \hline
			Project                       &    LC10\_021     &            AD509             &    17A-083     \\
			Configuration             & HBA\_DUAL\_INNER &              D               &       C        \\
			Pointing position                       &     $02^{\rm h}17^{\rm m}1.4^{\rm s}$, $+70^{\rm d}36^{\rm m}16.0^{\rm s}$ &    $02^{\rm h}17^{\rm m}0.0^{\rm s}$, $+70^{\rm d}36^{\rm m}36.0^{\rm s}$          &      $02^{\rm h}18^{\rm m}49.9^{\rm s}$, $+70^{\rm d}27^{\rm m}36.0^{\rm s}$ (E);                        \\
			(R.A., Dec.)                                      & & &  $02^{\rm h}14^{\rm m}31.0^{\rm s}$, $+70^{\rm d}41^{\rm m}04.0^{\rm s}$ (W)  \\
			Observing dates            &  2018 Sept. 26   & 2005 Dec. 4, Dec. 8, Dec. 12 &  2017 Jun. 3  \\
			Obs. IDs                   &     L670298      &              --              &       --       \\
			Calibrator(s)              &      3C~196      &        3C~286, 3C~352, 0217+738       &    3C 138, J0217+7349 \\
			Frequency (GHz)            &  $0.120-0.168$   &            $1.37-1.48$            &     $1.008-2.032$      \\
			Number of subbands  &       243        &              --              &      16       \\
			Bandwidth per subband (kHz)     &      195.3       &              --              &      64000       \\
			Channels per subband            &        16        &              --              &      64      \\
			On-source time (hr)        &        8         &             3.5              &        1.7 (E) + 1.7 (W)       \\
			Integration time (s)       &        1         &              13      &         5      \\
			Frequency resolution (kHz) &       12.2       &    12.5      &       1000         \\
			Correlations               &  XX, XY, YX, YY  &           RR, LL            & RR, RL, LR, LL \\
			Number of stations/antennas         &        60        &             23              &      27       \\ \hline
		\end{tabularx}\\
		\label{tab:obs} 
	\end{table*}
	
	% observations
	LOFAR High Band Antenna (HBA) observations of ClG0217 were performed for 8 hours on September 26, 2018 (project code: LC10\_021). The observational settings are identical to those used by the LOFAR Two-metre Sky Survey \citep[LoTSS;][]{Shimwell2017,Shimwell2019}. The 8-hour observation of the target field was bookended by 10-minute scans of calibrators (i.e. 3C~196 and 3C~48). The observations were performed with core and remote stations in the Netherlands and international stations in surrounding European countries.
	The distances between the Dutch stations that are used in this study range from 42~m to 115~km, corresponding to $20\,\lambda$ to $61\,k\lambda$ at the central frequency of 141~MHz. The high density of the LOFAR core stations makes it highly sensitive to diffuse sources. Details of the observations are given in Table~\ref{tab:obs}.
	
	% calibration: DiD, DD + beam correction
	The calibration of the LOFAR data was performed using the standard pipelines  (i.e. $\mathtt{Prefactor}$\footnote{\url{https://www.astron.nl/citt/prefactor}} and  $\mathtt{DDF-pipeline}$\footnote{\url{https://github.com/mhardcastle/ddf-pipeline}}) that have been developed for the LOFAR Surveys Key Science Project \citep{VanWeeren2016a,Shimwell2017,Shimwell2019,DeGasperin2019,Tasse2021}. The pipelines address the direction-dependent and direction-independent effects that are commonly known to affect low-frequency radio observations. Following the direction-dependent calibration, all sources outside a square region of $1\times1\,{\rm deg}^2$ centred on the cluster were subtracted from the $uv$ data. This subtracted data was then processed through several iterations of self-calibration to refine the image quality at the location of the target \citep{VanWeeren2020}. We note that the data was applied an approximate correction for the LOFAR primary beam attenuation by multiplication with a factor during the subtraction step. However, as the pointing position is at the location of the cluster, the factor is 1. Due to the large angular distance of the sources to the pointing centre of ClG0217 (i.e. up to 15\arcmin), the flux densities of these sources were not properly corrected. We applied the primary beam correction during the final imaging steps below. To enhance faint, diffuse emission in the final images, we applied Briggs' weightings of the $uv$ data using various \texttt{robust} values with \texttt{WSCLEAN} \citep[version 2.9.3;][]{Offringa2014,Offringa2017}. The imaging parameters are presented in Table \ref{tab:image_para}.We also used joined-channel and multi-scale deconvolution \citep{Offringa2017} with the \texttt{join-channels} (\texttt{channels-out}=6) and \texttt{multiscale} (\texttt{multiscale-scales}=[0, 3, 7, 25, 60, 150]) options. The built-in option \texttt{apply-primary-beam} was added in the \texttt{CLEAN} command to correct for the response of the LOFAR primary beam across the imaged region.
	
	% flux scale checking
	We examine the flux scale of the LOFAR images by comparing the flux densities of compact sources in our LOFAR images with those detected with the TIFR Giant Metrewave Radio Telescope (GMRT) 150~MHz Sky Survey-Alternative Data Release \citep[TGSS-ADR1;][]{Intema2017}. We selected compact sources that are brighter than 100~mJy and were detected with both the LOFAR and TGSS-ADR1 observations. The average flux densities for these sources in our LOFAR image are about 5 percent higher than those in the TGSS-ADR1 image. In this paper, we adopt a flux scale uncertainty of 10 percent on the LOFAR data as assumed by a number of previous studies \citep[e.g.][]{Bruno2021,Rajpurohit2021}.

	\begin{table*}
		\centering
		\caption{Imaging parameters.}
		\begin{tabularx}{\textwidth}{@{}Xcccccc@{}} 
			\hline\hline
			Data          &  $uv$-range  &       $\mathtt{Robust}^a$ &  $\mathtt{uvtaper}$    &         $\theta_\text{\tiny FWHM}$         &     $\sigma_{\text{\tiny rms}}$      &                  Figure                    \\
			& [k$\lambda$] &                            &  $[\arcsec]$ & [$\arcsec\times\arcsec$], ($PA^b[^\circ]$) & [$\upmu\text{Jy}\,\text{beam}^{-1}$] &                                                  \\ \hline 
			LOFAR         &  $0.12-61$   & 0       & 5  & $14.6\times10.1$ ($-85$)   & 160 & \ref{fig:lofar_hres}            \\
			&  $0.12-61$   & 0       & 15 & $26.6\times22.5$ ($29$)    & 240 & \ref{fig:label}    \\
			&  $0.12-61$   & 0.25    & 25 &  $45.7\times44.7$ ($52$)   & 335 & \ref{fig:df}            \\
			&  $0.12-61$   & 0       & 5  &  $16\times16^c$            & 190 &\ref{fig:spx}$^d$          \\
			&  $0.12-61$   & 0       & 30 &  $46\times46^c$            & 345 & \ref{fig:spx}$^d$, \ref{fig:spx_profile_halo}$^d$, \ref{fig:correlation}           \\
			&    $2-61$    & $-0.25$ & 5  &  $11.2\times7.5$ ($-89$)   & 140 & \ref{fig:ps}$^e$                \\ \hline
			VLA (C-array) &  $0.12-22$   & 0       & 8  &  $16\times16^c$            & 40  & \ref{fig:vla_images}, \ref{fig:spx}$^d$            \\
			VLA (D-array) &  $0.12-5.1$  & 0       & -- &  $46\times46^c$            & 85  & \ref{fig:vla_images}, \ref{fig:spx}$^d$, \ref{fig:spx_profile_halo}$^d$, \ref{fig:correlation}$^d$ \\
			&  $2.0-5.1$   & $-2$    & -- &  $28.4\times24.6$  ($51$)  & 80  & \ref{fig:ps}$^e$              \\ \hline
		\end{tabularx}\\	
		Notes: $^a$: Briggs weighting, $PA$: position angle with the reference axis to the N and counter-clockwise as positive, $^c$: smoothed, $^d$: spectral index, $^e$: beam uncorrected. 
		\label{tab:image_para}
	\end{table*}

	\subsection{VLA data}
	\label{sec:vla} 
	
	We observed ClG0217 with the VLA \textit{L}-band (1.5~GHz) in C array on June 3, 2017 (Project: 17A-083) and combine with archival data obtained with the VLA \textit{L}-band (1.4~GHz) D-array between December 4 and 12, 2005 (Project: AD509). Details of these observations are summarised in Table \ref{tab:obs}.
	
	The VLA \textit{L}-band C-array observations were performed using two pointings centred 13$\arcmin$ to the north-west and to the south-east of the centre of the cluster. This was done to increase the sensitivity to the sources in the peripheral regions of the cluster. We processed the data using the VLA $\mathtt{CASA}$ calibration pipeline version 4.7.2. The processing was optimised with the additional flagging and calibration to produce high-quality images. 3C 138 was used as amplitude and bandpass calibrator; and J0217+7349 was used for a phase calibration. After the initial imaging, several cycles of self-calibration were separately performed on the two pointings in order to refine the calibration. Finally, we make use of the \texttt{tclean} task in \texttt{CASA} \citep[version 5.6;][]{Mullin2007} to separately make intensity images for two pointings. A wideband correction for the primary beam attenuation using the \texttt{widebandpbcor} task was applied to the images. We then combined the images to make a mosaic of the cluster. The pixel values in the overlay regions between the pointings where the amplitude of the primary beams are larger than 0.2 were averaged, weighted by their amplitude.
	
	We reprocessed the VLA \textit{L}-band D-array data sets that were originally presented in \cite{Brown2011a}. The data was calibrated using $\mathtt{CASA}$ \citep[version 5.6;][]{Mullin2007}. The flux densities of the sources are calibrated using observations of 3C~286 and 3C~352 that we tied to the \cite{Perley2017} flux scale. The compact source 0217+738, which is located approximately 3~degree from ClG0217, was observed in-between target scans and was used as a phase calibrator. The calibration was performed for each data set in a standard manner (i.e. flagging of outliers, calibrating of the flux scale, antenna position, phase delay, bandpass, and complex gain). After the initial calibration, the data sets were combined and processed through multiple loops of self-calibration. The final, calibrated data was used to make an image of the cluster using Briggs' weighting (see Table \ref{tab:image_para}). The final image was also corrected for the attenuation of the primary beam (i.e. by using \texttt{pbcor=True} option in \texttt{tclean}). We adopted a flux scale uncertainty of 5 percent associated with the calibration of the VLA data. Due to the different settings (i.e. pointing positions) of the VLA C-array and D-array observations, we do not combine the data sets for final imaging.
	
	% source subtraction
	\subsection{Removal of discrete sources}
	\label{sec:subtract}
	
	The diffuse sources in ClG0217 are contaminated by discrete sources which we removed from the LOFAR and VLA D-array $uv$ data by subtracting their models. The removal of the discrete sources is essential for determining the flux density and spectra of the diffuse sources. To remove discrete sources, their source models at 141~MHz and 1.4~GHz were created by imaging the data at high-resolutions using only the baselines longer than 2~k$\lambda$ (i.e. corresponding to an angular size of smaller than $2.1\arcmin$ or $382\,{\rm kpc}$ at $z=0.18$) and applying Briggs' weighting to give more weight to the long baselines (i.e. $\texttt{robust}=-0.25$ for LOFAR and $\texttt{robust}=-2$ for VLA). An advantage of the combination of these settings is that they prevented any significant large-scale emission from being present in the model images which is subtracted from the data. Moreover, when including more data for imaging, the noise level of the model images is lowered which allows the detection of faint, compact sources. In Appendix \ref{sec:app_ps}, we present the LOFAR and VLA high-resolution images that are obtained with these tuned imaging parameters. %The LOFAR and VLA high-resolution images (Figures\,\ref{fig:lofar_hres} and \ref{fig:vla_images}) that are obtained with these tuned imaging parameters show just the compact sources in the field of ClG0217, but the diffuse emission from ClG0217 is not clearly seen. 
	The discrete sources are identified with \texttt{PyBDSF} using a source detection threshold of $4\sigma$ (i.e. $\texttt{thresh\_pix}=4$). These clean-component models of the contaminating sources were then subtracted from the LOFAR and VLA $uv$ data.
	
	\subsection{Spectral index maps}
	\label{sec:spec}
	
	We combined the LOFAR 141~MHz data with the VLA 1.4~GHz (D-array) and 1.5~GHz (C-array) data to make spectral index maps of the radio sources in the cluster field. When making the LOFAR and VLA intensity images, we selected the $uv$ data with a common inner $uv$ cut for both data sets. Here we used an inner $uv$ cut of $0.12\,k\lambda$ for the $16\arcsec$ and $46\arcsec$ resolution maps. Briggs' weighting of the $uv$ data ($\texttt{robust}=0$) is combined with tapering of the outer baselines ($\texttt{uvtaper}$) to obtain images with the resolutions that are close to the targeted resolutions (i.e. $16\arcsec$ and $46\arcsec$). The imaging of the LOFAR and VLA data were done with $\mathtt{WSClean}$ and $\mathtt{CASA}$, respectively (see Section \ref{sec:lofar} and \ref{sec:vla}). The LOFAR and VLA images were smoothed to the final resolutions, aligned using compact sources, and regridded to the same pixel size. The properties of the final images are given in Table \ref{tab:image_para}. Spectral index maps were made with the LOFAR and VLA final images, following the power-law relation of the synchrotron spectrum ($S\propto\nu^\alpha$). The spectral index errors are added in quadrature from the flux scale uncertainty and the image noise.
	
	\subsection{Chandra data}
	\label{sec:chandra}
	
	The \emph{Chandra} ACIS-I observations of ClG0217 were performed for 25~ks on November 28, 2014 (ObsID: 16293). The calibration of the data was done in \cite{Zhang2020}. For completeness, we briefly describe it here. The level-2 event files were obtained with the \emph{Chandra} Interactive Analysis Observations (CIAO)\footnote{\url{https://cxc.harvard.edu/ciao}} package using the \texttt{chandra\_repro} task. The X-ray SB image in the energy band $1-3$~keV was extracted with the \texttt{fluximage} task. The SB image was then subtracted the non X-ray background (NXB), corrected for the vignetting, and adaptively smoothed.
	
	%%%%%%%%%%%%%%%%%%%%%%%%%%%%%%%%%%%%%%%
	\section{Results and analysis}
	\label{sec:res}
	
	\begin{figure*}[!h]
		\centering
		\includegraphics[width=0.6\textwidth]{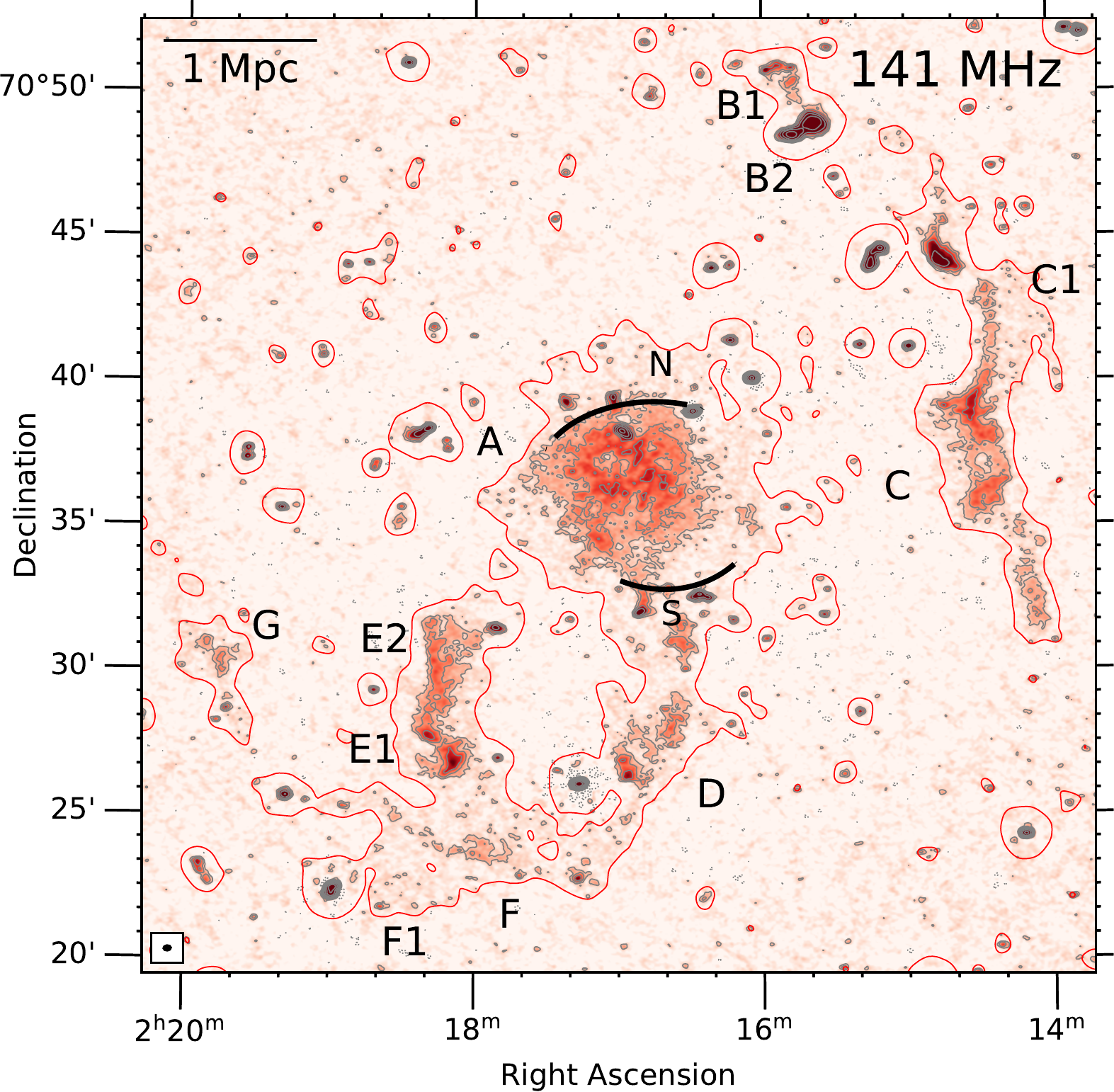}
		\caption{LOFAR $14.6\arcsec\times10.1\arcsec$-resolution image of the galaxy cluster ClG0217. The image shows the roundish halo emission (A) at the cluster centre and relics (C, D, F and G) in the outskirts. The first grey contour starts from $3\sigma$, where $\sigma=160\,\upmu{\rm Jy\,beam}^{-1}$. The subsequent contours are multiplied by a factor of 2.  The red contour level is $3\sigma$, where $\sigma=335\,\upmu{\rm Jy\,beam}^{-1}$ with the beam size of $45.7\arcsec\times44.7\arcsec$. The synthesised beam of $14.6\arcsec\times10.1\arcsec$ is shown in the bottom left corner. The arc lines show the locations of detected X-ray SB discontinuities in the northern and southern edges of the halo that are found in \cite{Zhang2020}.}
		\label{fig:lofar_hres}
	\end{figure*}

	\begin{table*}
		\centering
		\caption{Radio source properties}
		\begin{tabularx}{\textwidth}{@{}XYYYYYY@{}}
			\hline\hline
			Source  & $S_{\rm 141\,MHz}^a$ &        $P_{\rm 141\,MHz}^b$         & $S_{\rm 1.4\,GHz}^a$ &        $P_{\rm 1.4\,GHz}^b$         &    $\alpha$     & LAS$^c$ \\
			&        [mJy]         & [$10^{24}\,{\rm W}\,{\rm Hz}^{-1}$] &        [mJy]         & [$10^{23}\,{\rm W}\,{\rm Hz}^{-1}$] &                 &  [Mpc]  \\ \hline
			A       &    $613.4\pm61.5$    &        $56.4\pm5.7$                 &     $51.1\pm2.7$     &            $47.0\pm2.5$             & $-1.07\pm0.05$  &   1.8   \\
			&       $623.6\pm62.7^d$           &       $56.9\pm5.7^d$                              &    $58.3\pm3.4^d$                  &              $53.2\pm3.1^d$                       &         $-1.02\pm0.05^d$        &   --      \\
			B1      &     $24.6\pm2.6$     &            $2.1\pm0.2$             &    $5.2\pm0.3$      &            $4.5\pm0.3$            & $-0.67\pm0.05$  &   0.3   \\
			B2      &     $170.3\pm17.1$     &             $15.1\pm1.5$             &     $23.5\pm1.2$      &             $2.1\pm0.1$             & $-0.86\pm0.05$  &   --    \\
			C  &    $211.4\pm21.3$    &            $19.3\pm2.0$             &     $20.3\pm1.1$     &            $18.5\pm1.0$             & $-1.01\pm0.05$  &   2.3   \\
			C1  &    $13.7\pm1.8$    &           $1.3\pm0.2$            &     $<0.3^c$     &            $<0.3$             & $<-1.7$  &   0.7   \\
			D       &    $127.0\pm13.0$    &            $12.3\pm1.3$             &     $5.7\pm0.5$      &            $5.5\pm0.5$           & $-1.34\pm0.06$  &   1.8   \\
			E1      &     $84.2\pm8.5$     &             $7.6\pm0.8$            &     $9.7\pm0.6$      &             $8.7\pm0.5$            &  $-0.93\pm0.05$  &   0.5   \\
			E2      &     $93.3\pm9.5$     &             $8.8\pm0.9$             &     $5.9\pm0.4$      &             $5.5\pm0.4$            &  $-1.19\pm0.05$  &   1.0   \\
			F       &     $59.8\pm6.3$     &             $5.5\pm0.6$             &     $4.8\pm0.4$      &             $4.4\pm0.4$             & $-1.09\pm0.06$  &   1.7   \\
			F1   &     $15.1\pm1.8$     &             $1.6\pm0.2$             &     $<0.3^c$      &             $<0.3$             & $<-1.7$  &   0.6   \\
			G       &     $31.4\pm3.5$     &             $2.9\pm0.3$             &     $3.1\pm0.3$      &             $2.8\pm0.3$             & $-1.00\pm0.06$  &   1.6   \\ \hline
		\end{tabularx}\\
		Notes: %The regions where the flux density is extracted are shown in Figure~\ref{fig:label}. 
		$^a$: the flux density is calculated for all pixels that are detected above $2\sigma$, except in the case of $^d$ where flux density is obtained by fitting the surface brightness of the halo with a 2-dimensional circular model. $^b$: k-corrected radio power, $P=4\pi D^2_L S_\nu / (1+z)^{(1+\alpha)}$, where $D_L$ is the luminosity distance. $^c$: Largest Angular Size. $^c$: upper limit estimated by $\sigma\times\sqrt{N}$, where $N$ is the source area in beam unit.   
		\label{tab:source_properties} 
	\end{table*}

	\begin{figure*}
		\centering
		\includegraphics[width=0.45\textwidth]{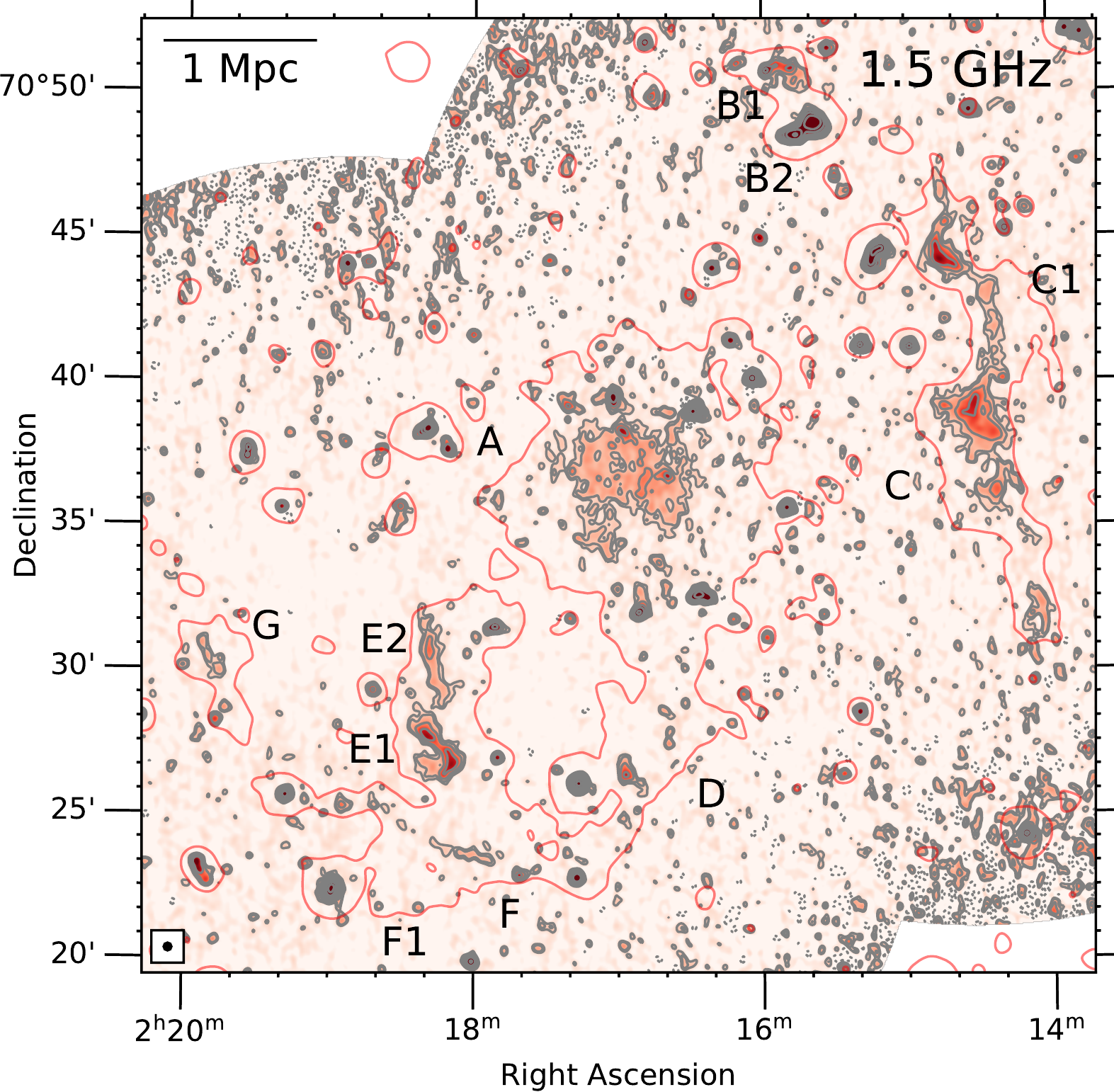} \hfil
		\includegraphics[width=0.45\textwidth]{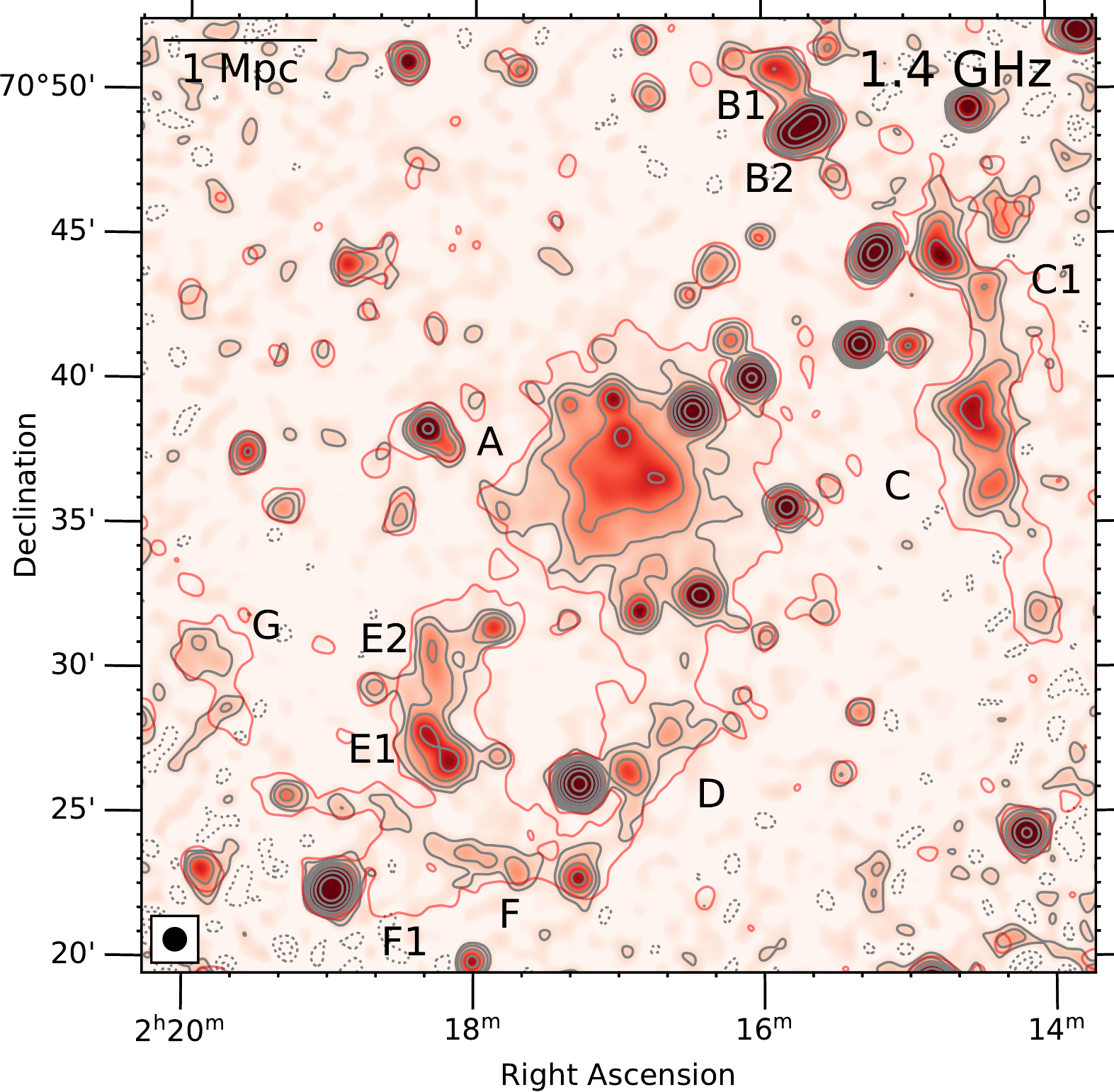}
		\caption{VLA C (\textit{left}) and D-array (\textit{right}) images. Contour starts from $3\sigma$, where $\sigma_{\rm C-array}=40\,\upmu{\rm Jy\,beam}^{-1}$ and $\sigma_{\rm D-array}=85\,\upmu{\rm Jy\,beam}^{-1}$. The subsequent contours are multiplied by a factor of 2. The synthesised beams are drawn in the bottom left corners (i.e. $16\arcsec\times16\arcsec$ for the C-array image and $46\arcsec\times46\arcsec$ for the D-array image). The LOFAR $45.7\arcsec\times44.7\arcsec$-resolution contour is the same as that in Figure~\ref{fig:lofar_hres}. 
		}
		\label{fig:vla_images}
	\end{figure*}

	\begin{figure}[!h]
		\centering
		\includegraphics[width=1\columnwidth]{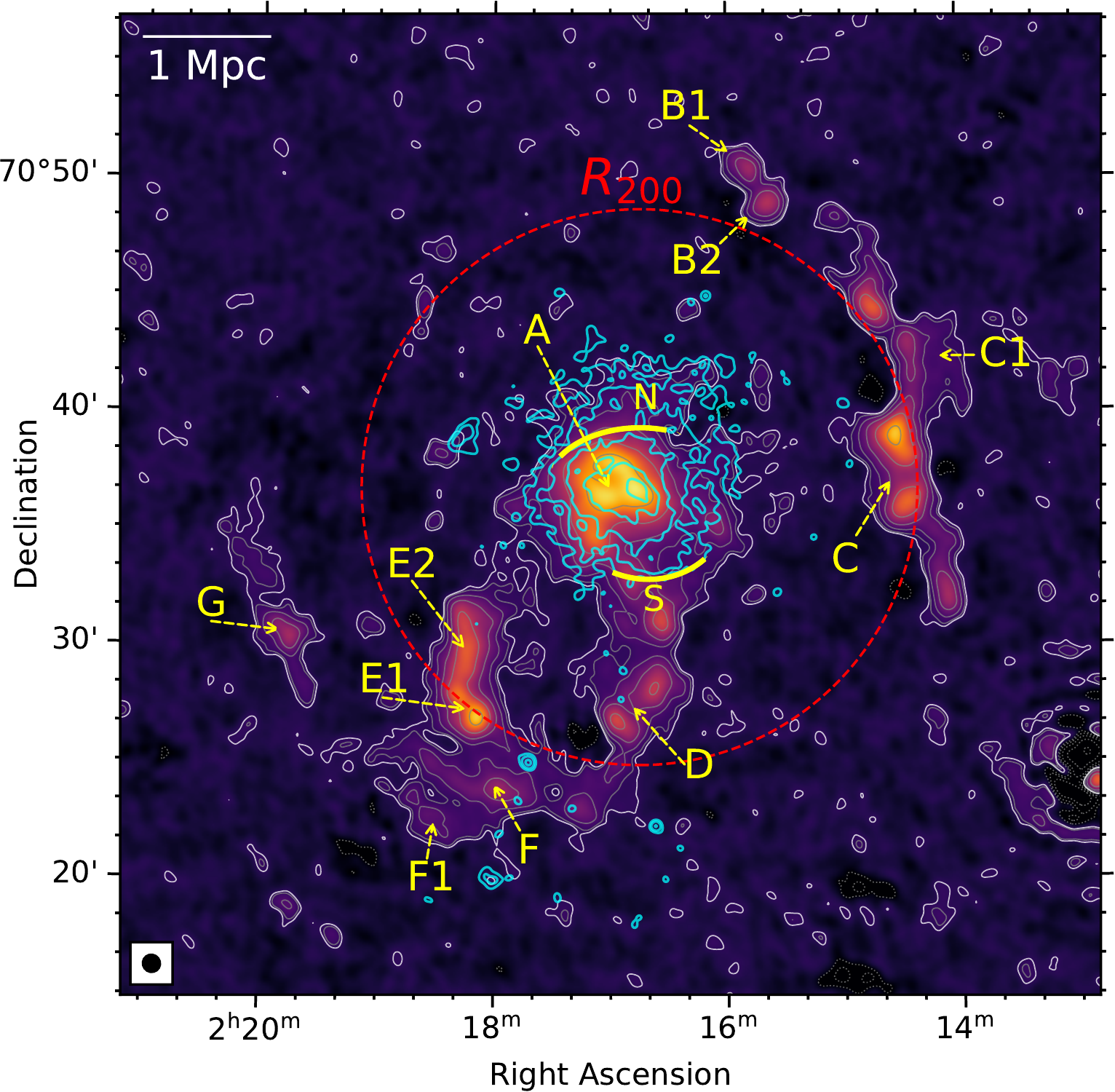}
		\caption{LOFAR $45.7\arcsec\times44.7\arcsec$-resolution image of ClG0217. The contaminating discrete sources are removed from the data. The LOFAR grey and \textit{Chandra} cyan contours are drawn at $\pm[1,2,4,8,16]\times3\sigma$, where $\sigma_{\rm LOFAR}=335\,\upmu{\rm Jy\,beam}^{-1}$ and $\sigma_{{\rm \textit{Chandra}}}=1.3\times10^{-6}{\rm cts\,arcmin^{-2}cm^{-2}}$. The LOFAR white contour is drawn at $2\sigma_{\rm LOFAR}$. The \textit{Chandra} data is smoothed with a Gaussian beam of 5-pixel sigma where pixel size is $1.968\arcsec$. The red dashed circle has a radius of $R_{200} = 2.3\,{\rm Mpc}$. The arc lines show the locations of the northern and southern X-ray SB discontinuities.}
		\label{fig:df}
	\end{figure}

	\begin{figure}[!h]
		\centering
		\includegraphics[width=0.8\columnwidth]{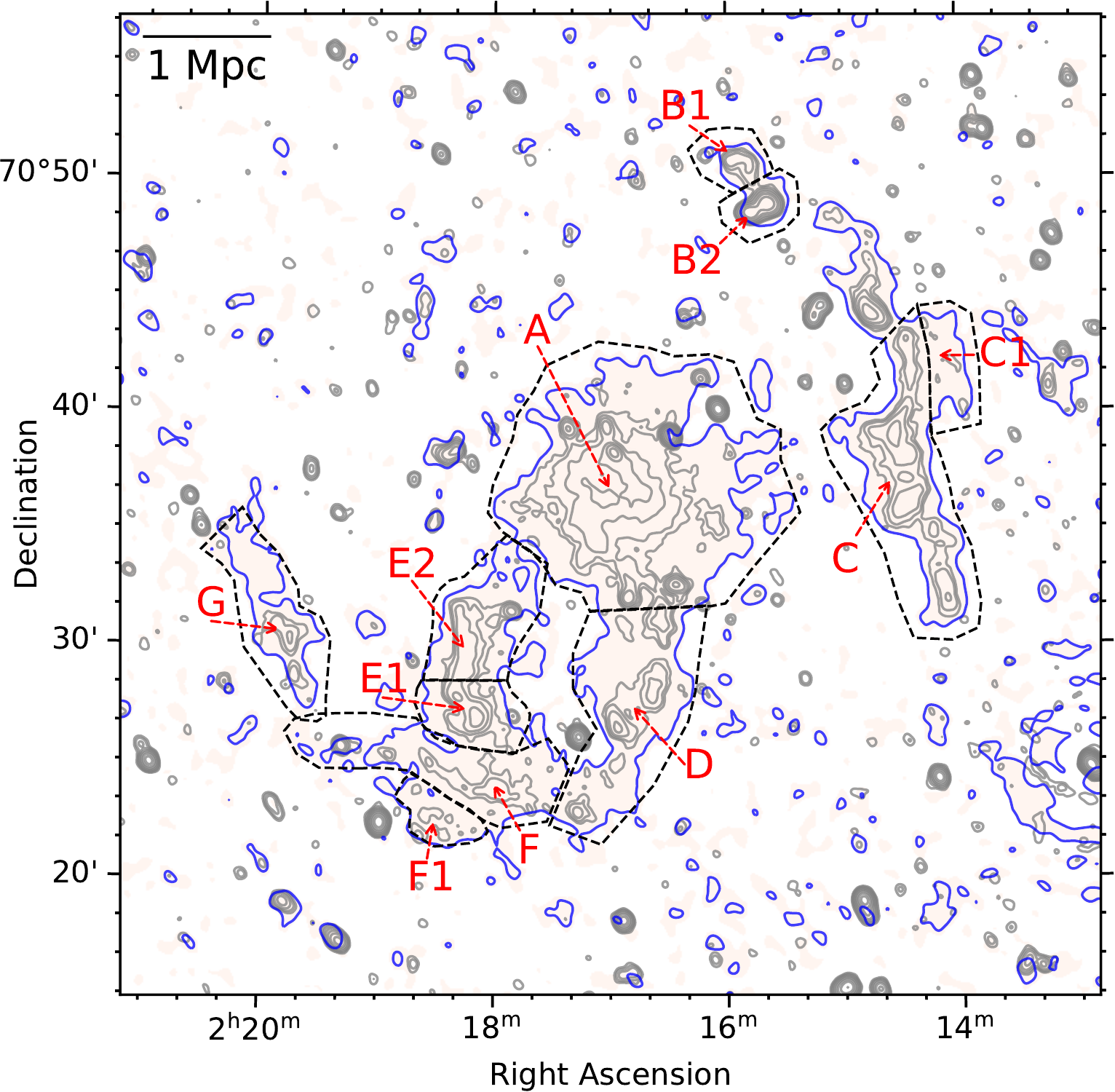} 
		\caption{LOFAR $26.6\arcsec\times22.5\arcsec$-resolution image. The contour levels are the same as those in Figure~\ref{fig:lofar_hres} with $\sigma=240\,\upmu{\rm Jy\,beam}^{-1}$. The compact-source subtracted blue contour is drawn at $2\sigma$, where $\sigma=335\,\upmu{\rm Jy\,beam}^{-1}$ and the beam size of $45.7\arcsec\times44.7\arcsec$. The black dashed lines show the regions used to measure the flux densities from the cluster diffuse sources and the measurements are summarised in Table~\ref{tab:source_properties}.
		}
		\label{fig:label}
	\end{figure}
	%In Figure~ \ref{fig:df} we present the LOFAR 141~MHz image of the galaxy cluster ClG0217 that has discrete sources removed (see Section \ref{sec:subtract}). The LOFAR data confirms the presence of diffuse radio emission in the cluster central region (i.e. source A) and elongated radio sources (i.e. sources B1, B2, C, D, E1, E2, F, and G) in the outskirts that were detected at 1.4~GHz with the VLA observations by \cite{Brown2011a}. We roughly separate the diffuse radio sources with dashed lines as shown in Figure~\ref{fig:flux_regions}.
	
	In Figure~\ref{fig:lofar_hres}, we present our deep LOFAR 141~MHz $14.6\arcsec\times10.1\arcsec$-resolution image of the galaxy cluster ClG0217. For the first time, the LOFAR observations allowed us to image the cluster at a frequency below 350\,MHz. 
	
	We have labelled the sources following \cite{Brown2011a} and split sources B and E in \cite{Brown2011a} into separated sources B1+B2 and E1+E2. We also label sources C1 and F1 that are newly detected with the LOFAR observations. The VLA C- (1.5~GHz) and D-array (1.4~GHz) images are shown in Figure~\ref{fig:vla_images}. The LOFAR~141~MHz and VLA~1.5~GHz images confirm the presence of central halo emission (A) and several elongated diffuse radio sources (C, D, E1+E2, F, and G) in the outskirts, detected previously at 1.4~GHz by \cite{Brown2011a}. In the subsections below, we present the flux density and spectral measurements of the diffuse sources in the cluster and summarise the results in Table \ref{tab:source_properties}.
	
	\subsection{Central radio halo}
	\label{sec:res_halo}
	
	% morphology, location, size
	As shown in Figures~\ref{fig:lofar_hres} and \ref{fig:vla_images}, the radio halo has a fairly round shape. The largest linear size (projected) of the halo is about 1.8~Mpc and 1.5~Mpc at 141~MHz and 1.4\,GHz (measured from the D-array image), respectively. Apparently, the sensitive low-frequency observations allow us to reveal more extended very low surface brightness emission. In the new VLA C-array image in Figure \ref{fig:vla_images} (\textit{left}), the halo size is significantly smaller, i.e. 800~kpc. This is likely due to the fact that the VLA C-array observations lack the sensitivity (i.e. signal to noise ratio) to the large-scale (i.e. $\sim$~$10\arcmin$ of the halo) emission although%, in principle, %the short baseline of $120\lambda$ used in the imaging is capable of mapping diffuse emission of scale up to $35\arcmin$ and 
	the largest angular detectable scale of the C array\footnote{\url{https://science.nrao.edu/facilities/vla/docs/manuals/oss2017A/performance/resolution}} is $16\arcmin$. 
	
	The morphology of the halo emission is similar at 141\,MHz and 1.4\,GHz. Figure~\ref{fig:lofar_hres} shows some substructures within the halo emission at 141\,MHz. Multiple unrelated discrete sources are also embedded in the southern and northern regions of the halo.
	
	The low-resolution ($45.7\arcsec\times44.7\arcsec$), point-source subtracted LOFAR image of the halo superimposed on the \textit{Chandra} X-ray contours is shown in Figure~\ref{fig:df}. As typical for halos, the radio emission from the halo seems to follow the thermal X-ray emission, implying a connection between the thermal and non-thermal components of the ICM.
	
	Interestingly, Figure~\ref{fig:df} also shows that the southern and eastern regions of the halo are apparently connected to elongated diffuse sources D and E2, respectively. The connection between the halo and sources D and E2 can only be seen at 141~MHz, suggesting that these regions have steep spectra ($\alpha<-1.6$).
	
	\subsubsection{Flux density and radio power}
	\label{sec:res_flux}
	
	% flux density and integrated spectral measurements
	The 141~MHz flux density of the radio halo measured within the $3\sigma$ contour of the discrete-source subtracted image is $597.7\pm59.9$~mJy. In Figure \ref{fig:label}, we show the low-resolution image and the regions used for extracting the halo flux density. The measured halo flux density increases by 2.6 percent to $613.4\pm61.5$~mJy when all pixels above $2\sigma$ are included. At 1.4~GHz, we measure the flux density for the halo integrating over the $3\sigma$ and $2\sigma$ regions to be $46.9\pm2.5$~mJy and $51.1\pm2.7$~mJy (a 9.0 percent increase), respectively. Using the same VLA data sets, \cite{Brown2011a} reported a higher value (i.e. $58.6\pm0.9$~mJy or $\sim$15 percent higher than our $2\sigma$ measurement). It is unclear where this difference comes from. It could be due to the differences in the thresholds and/or the absolute flux scale that are not described in \cite{Brown2011a}. The deconvolution algorithms might also contribute to the difference in the flux density reported in this work and  \cite{Brown2011a}.
	
	% fitting
	The application of a threshold cut to the pixel values used when integrating the halo flux might bias the measurement low, due to missing flux from the faint emission in the outer regions of the radio halo. For comparison and better recovery of the low-SB emission, we fit the SB of the halo with a 2-dimensional circular exponential model of the form,
	\begin{equation}\label{eq:I}
		I (r)=I_0\exp(-r/r_e),
	\end{equation}
	where the fitting parameters $I_0$ and $r_e$ are the SB at the halo centre and the \textit{e}-folding radius, respectively \citep{Murgia2009}. We made use of the \texttt{Halo-FDCA}\footnote{\url{https://github.com/JortBox/Halo-FDCA}} code \citep{Boxelaar2021} that has been used to estimate the flux densities of the radio halos detected in the HETDEX field of the LoTSS survey \citep{VanWeeren2020}. The code employs a Markov Chain Monte--Carlo method to search for the best-fit parameters and the associated uncertainties. In the halo of ClG0217, we find the best-fit parameters to be $I_0=15.51\pm0.18\,{\rm \upmu Jy\,arcsec^{-2}}$ and $r_e=255\pm2\,{\rm kpc}$ at 141~MHz and $I_0=1.19\pm0.04\,{\rm \upmu Jy\,arcsec^{-2}}$ and $r_e=281\pm7\,{\rm kpc}$ at 1.4~GHz. The corner plots showing the constraints of the fitting are presented in Appendix~\ref{sec:app_fitting}. The $e$--folding radius at 1.4~GHz is larger than that at 141~MHz, indicating the full extent is not detected with the current shallow data. Alternatively, the fitting might be biased by possible excess radio emission in the outer regions of the halo at 1.4~GHz that is associated with X-ray SB discontinuities to the northern, southern and north-eastern regions (discussed in Section \ref{sec:res_NE}).
	
	Using the best fit parameters, the flux density is integrated to a defined radius. %For ClG0217, the flux density is integrated within a radius of $4r_e$. This consists of 90 percent of the flux density when integrating the flux density to infinity. 
	The flux density for the halo in ClG2017 is $623.6\pm62.7$~mJy at 141~MHz and $58.3\pm3.4$~mJy at 1.4~GHz. The flux density above is integrated up to a radius of $4r_e$ which is slightly larger than the half of the halo size  (i.e. $900$~kpc) within the $2\sigma$ contours. This consists of 90 percent of the flux density when integrating to infinity. In order to obtain the total errors in the flux density, the uncertainties associated with the model fitting and the LOFAR and VLA flux calibration (i.e. $10\%$ and $5\%$, respectively) are added in quadrature. The corresponding \textit{k}-corrected radio power for the radio halo is $56.9\pm5.7\times10^{24}\,{\rm W\,Hz^{-1}}$ at 141~MHz and is $53.2\pm3.1\times10^{23}\,{\rm W\,Hz^{-1}}$ at 1.4~GHz. 
	
\begin{figure}
		\centering
		\includegraphics[width=1\columnwidth]{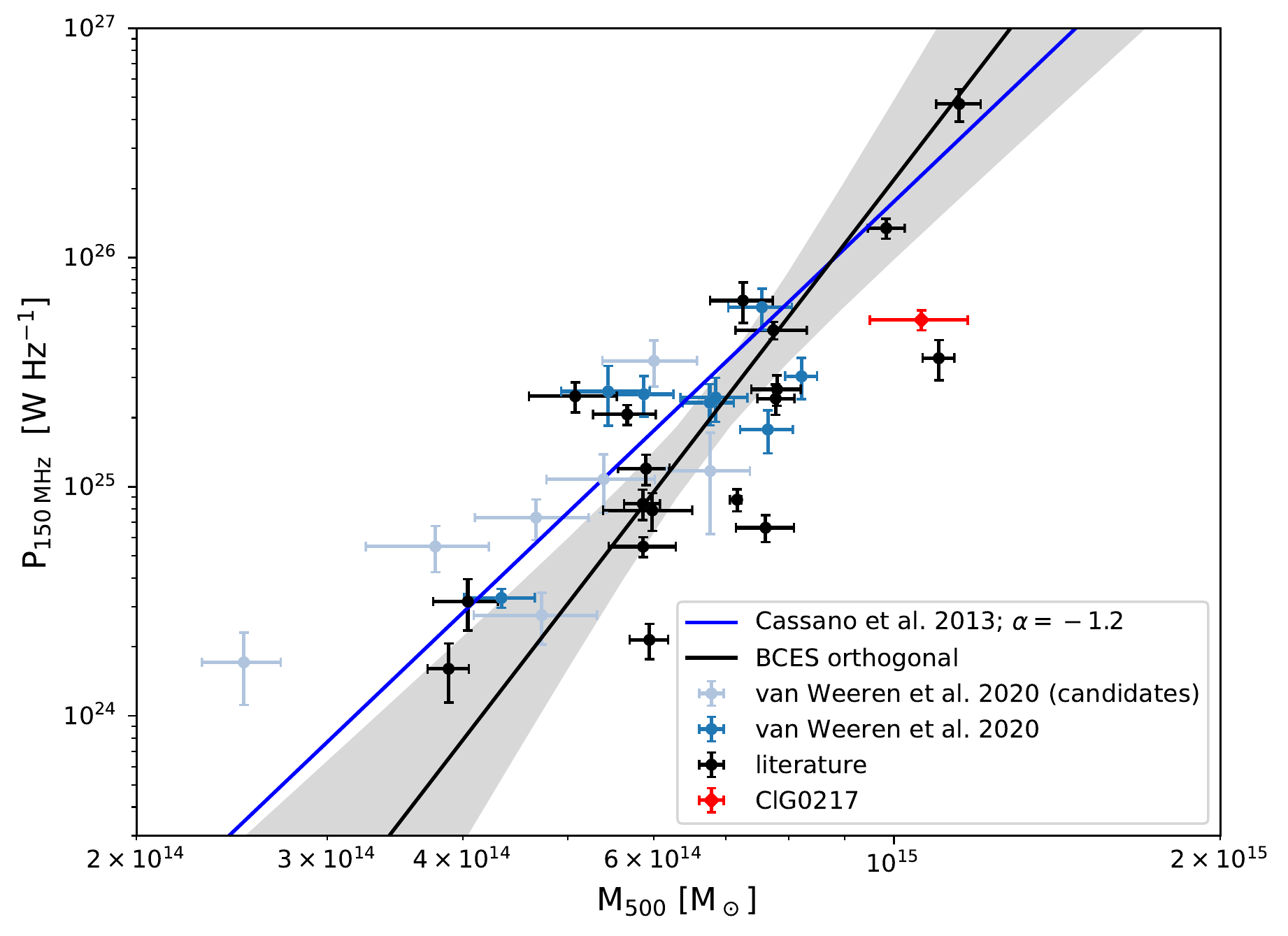}
		\caption{The $P_{\rm 150\,MHz}-M_{500}$ correlation including the data point for the radio halo in ClG0217 (in red). The plot is adapted from \cite{VanWeeren2020}. 
		}
		\label{fig:PM}
	\end{figure}

	Finally, in Figure~\ref{fig:PM}, we show the radio power versus cluster mass relation of known halos. The radio power of the ClG0217 halo is a bit lower than expected from the correlation between radio power and cluster mass, even though one should note the large scatter of the data points around the scaling relation. 
	
	\subsubsection{Spectral index measurements}
	\label{sec:res_spx}
	
	\begin{figure*}
		\centering
		\includegraphics[width=1\textwidth]{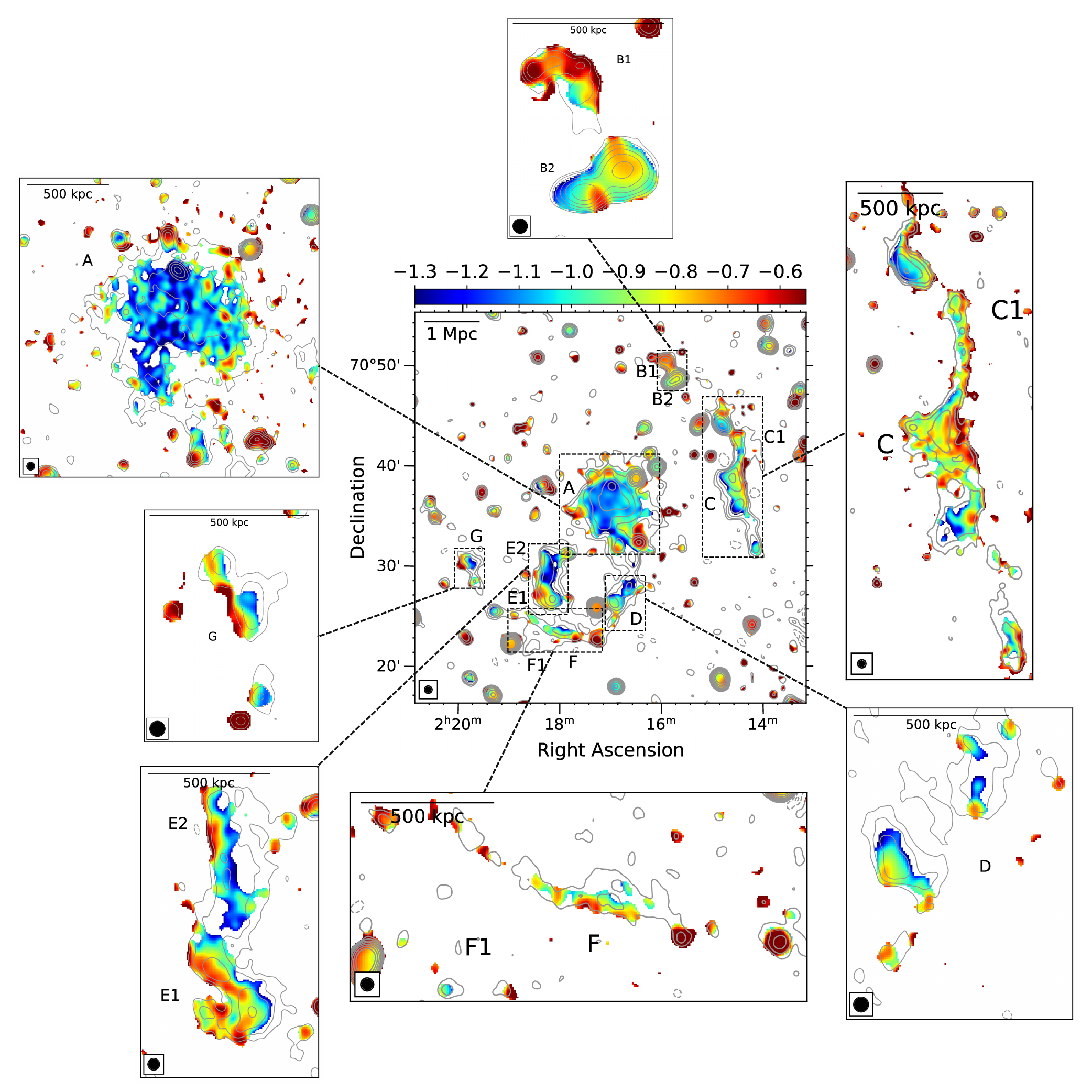}	%
		\caption{Spectral index map (central) between $141\,{\rm MHz}$ and $1.4\,{\rm GHz}$ at the resolution of $46\arcsec$. The cut-out images show the distribution of spectral indices between $141\,{\rm MHz}$ and $1.5\,{\rm GHz}$ at $16\arcsec$ resolution. The LOFAR contour levels are $\pm[1,2,4,8,16] \times3\sigma$, where $\sigma=345\,\upmu{\rm Jy\,beam}^{-1}$ (beam$_{\rm FWHM}=46\arcsec$) and $\sigma=190\,\upmu{\rm Jy\,beam}^{-1}$ (beam$_{\rm FWHM}=16\arcsec$) for the central and the cut-out images, respectively. The corresponding maps for the spectral index errors are presented in Appendix~\ref{sec:app_spx_err}.}
		\label{fig:spx}
	\end{figure*}
	
	% spatial spectral index analysis
	Using the flux density estimates obtained from the exponential model fitting, we find that the integrated spectral index between 141~MHz and 1.4~GHz for the radio halo is $-1.02\pm0.05$, which is consistent with the value of $-1.07\pm0.05$ that is obtained using the measurements from the $2\sigma$ threshold cut method. When combining our measurements with the 325~MHz flux density reported in \cite{Brown2011a}, we obtain integrated spectral indices of $-0.78\pm0.16$ between 141~MHz and 325~MHz and $-1.16\pm0.07$ between 325~MHz and 1.4~GHz. This implies that the radio halo has a curved spectrum with steeper index at high frequencies. However, we also note that the 325~MHz observations have a resolution too low to resolve the compact sources to the northern edge of the halo, as seen in Figure 2 of \cite{Brown2011a}. These compact sources might contaminate the flux measurement at 325~MHz, making the halo spectra deviate from its true values. We estimate that the halo follows a simple power-law spectrum if 20~percent of its 325~MHz flux density is subtracted due to the possible contamination by the compact sources.
	
	\begin{figure*}
		\centering
		\includegraphics[width=0.45\textwidth]{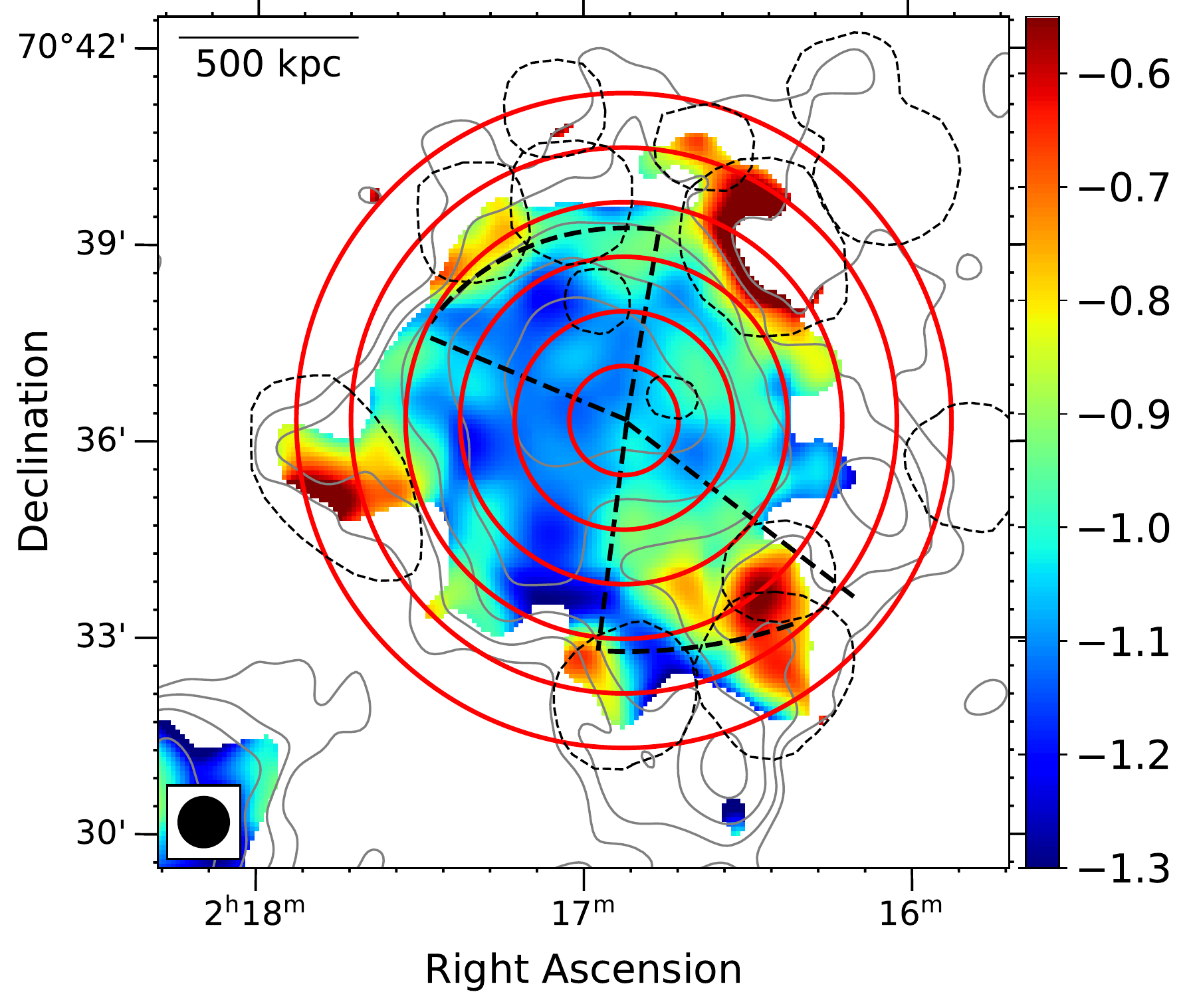} \hfil
		\includegraphics[width=0.54\textwidth]{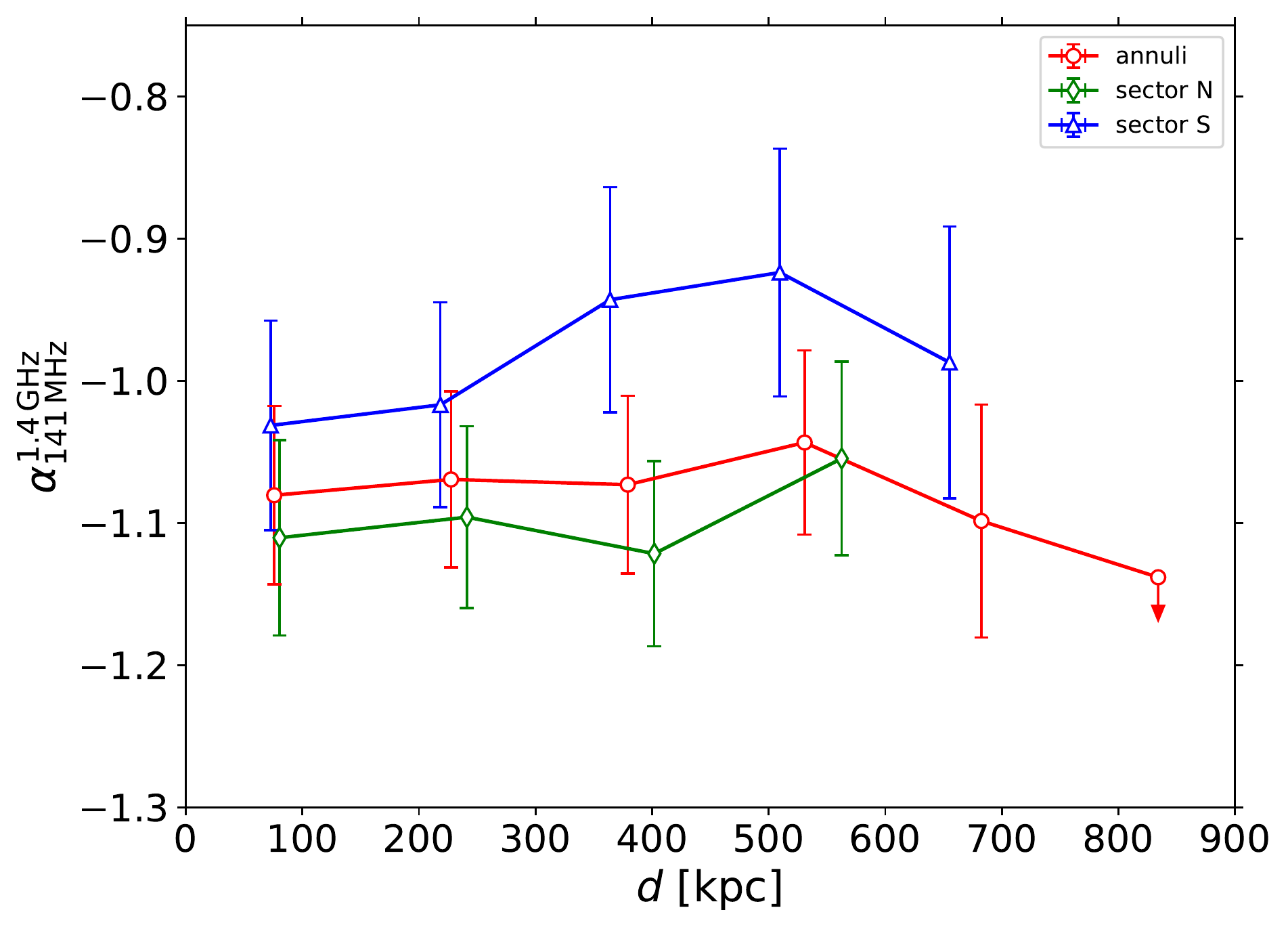}
		\caption{\textit{Left}: 141~MHz -- 1.4~GHz spectral index map in the halo region. The discrete sources are removed from the data before making the spectral index map. The red annuli and black dashed sectors are where the spectral indices are estimated and shown in the \textit{right} panel. The outermost regions of the N and S sectors are where X-ray SB discontinuities are found in \cite{Zhang2020}. The thin dashed lines are the regions where discrete sources are subtracted or show locally flat spectra which are excluded from the profiles. The LOFAR grey contours are $\pm[1,2,4,8,16] \times3\sigma$, where $\sigma=345\,\upmu{\rm Jy\,beam}^{-1}$ (beam$_{\rm FWHM}=46\arcsec$). \textit{Right}: The spectral index profiles from the halo centre towards the outskirts and towards the N and S sectors are plotted.}
		%[width: S: 53 (5 points), N: 48  (4 points), anulli: 50 (6 points). replacing left image]
		\label{fig:spx_profile_halo}
	\end{figure*}
	
	In Figure~\ref{fig:spx}, we show the low-resolution ($46\arcsec$) spectral index map of ClG0217 that was made with the LOFAR 141~MHz and VLA 1.4~GHz data. The cut-out images show the distribution of the spectral index in the source regions at a resolution of $16\arcsec$. The corresponding spectral index error maps are presented in Appendix~\ref{sec:app_spx_err}. In the central region of the halo, the spectral index remains roughly constant, but it seems to flatten in the outer regions. We note that the spectral index in the outermost regions of the halo might be contaminated by the imperfection of the subtraction of discrete sources. 
	
	To examine the radial spectral distribution of the halo, we extract the spectral index in circular annuli that have a width of $50\arcsec$ (i.e. slightly larger than the beam size of $46\arcsec$). 
	These regions are shown in the \textit{left} panel of Figure~\ref{fig:spx_profile_halo}. We note that, as mentioned in Section \ref{sec:subtract}, the discrete sources were subtracted from both LOFAR and VLA images. The resulting radial spectral index profile that is shown in the \textit{right} panel of Figure~\ref{fig:spx_profile_halo} suggests a uniform spectral index of $-1.07\pm0.02$ within a radius of 680~kpc (i.e. over an area of $1.5\,{\rm Mpc^2}$). Although there is hint of radial spectral flattening within the radius of $\sim$600~kpc, the spectral trend is within the $1\sigma$ uncertainty. In the outer region ($r\geq830\,{\rm kpc}$), the index becomes steeper than $-1.14$. We note that deeper radio observations are required to estimate the spectral index as the low surface brightness emission in this region may not be fully deconvolved. The errors estimated in the profile do not include the uncertainty associated with the flux scale calibration as, if included, the spectral index is systematically rather than randomly impacted. However, if the flux scale uncertainty is added in quadrature with the image noise, the typical errors for the spectral indices are $0.06$. Hence, this does not change our conclusion on the spectral index uniformity in the central region of the radio halo.
	
	\subsubsection{The north-eastern edge}
	\label{sec:res_NE}
	
\begin{figure}
		\centering
		\includegraphics[width=1\columnwidth]{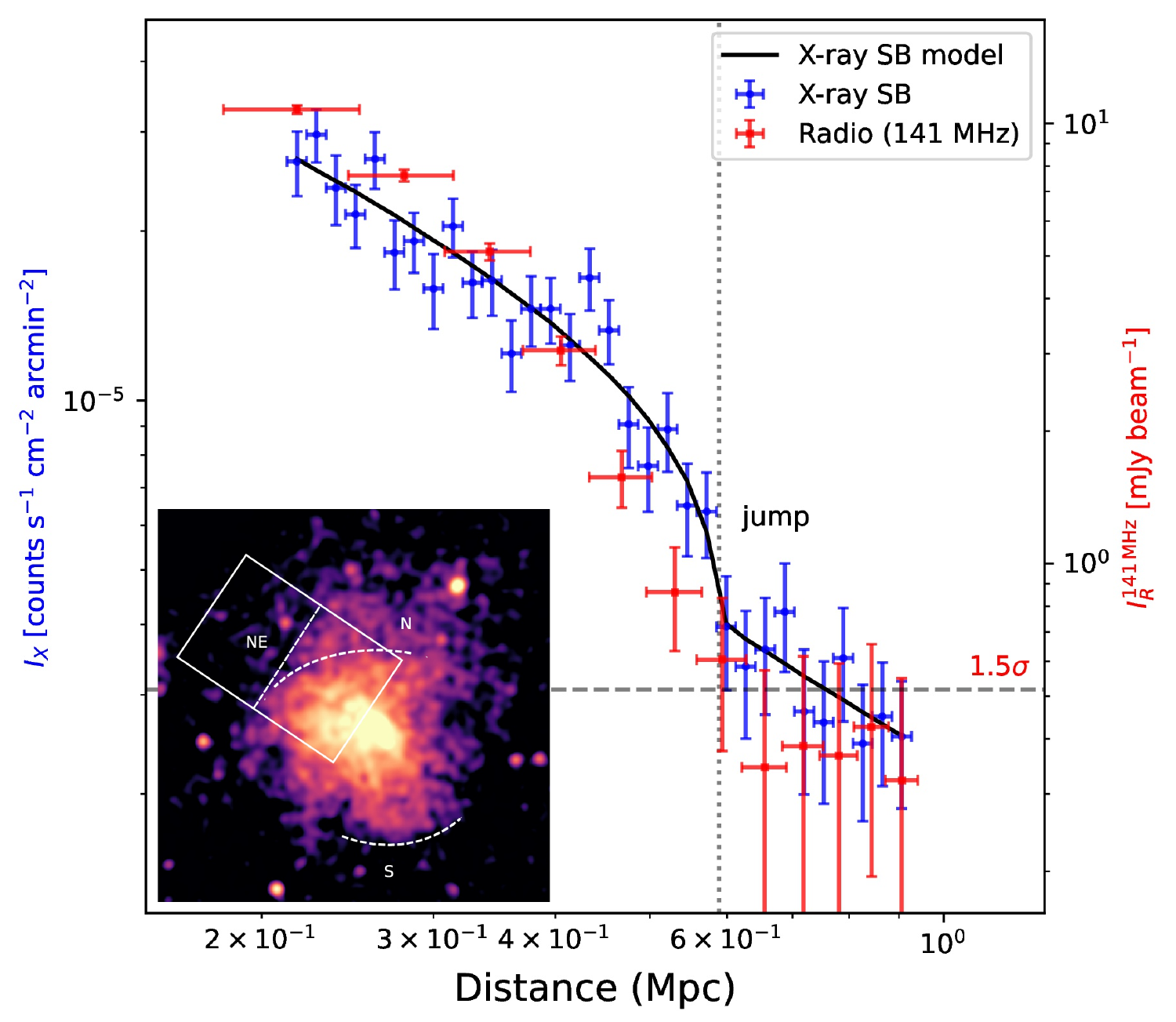}	%
		\caption{X-ray and radio (141~MHz) SB profiles towards the north-east direction of the halo. The X-ray SB model indicates a discontinuity marked by the vertical dotted line. In the overlay \textit{Chandra} image, the extracted north-eastern region is shown with the rectangles. The dashed N and S arcs are the X-ray SB edges found by \cite{Zhang2020}.}
		\label{fig:NE_edge}
	\end{figure}

	A recent study by \cite{Zhang2020} found discontinuities in the X-ray SB towards the northern and southern edges of the radio halo in the VLA 1.4~GHz data. The locations of these discontinuities are also shown in Figure \ref{fig:df}. In addition to these two edges, we note that the north-eastern edge of the radio halo in the LOFAR 141~MHz map could be associated with a new X-ray SB discontinuity. To examine this possibility, we extract a \emph{Chandra} SB profile using the box regions defined in the inset of Figure~\ref{fig:NE_edge}. The resulting profile is shown in Figure~\ref{fig:NE_edge}. We fit the X-ray data using a projected broken power-law density model \citep{Owers2009} and obtained a best-fit density jump of $1.92\pm0.34$ at the radius of $195\arcsec\pm8\arcsec$. Meanwhile, we extract the radio SB profile in this region from the $46\arcsec$-resolution map. The radio profile nicely follows the trend of the X-ray SB before the jump and drops quickly at the X-ray SB jump. The five outermost radio data points below $1.5\sigma_\mathrm{rms}$ are regarded as the radio background. The new north-eastern X-ray edge partly overlaps with the previously reported northern edge in \cite{Zhang2020}.
	
	Although the nature of the X-ray SB discontinuities is still unknown and the relation between the north and the north-eastern discontinuities is unclear yet, if they have a nature of a shock front, they might affect the spectra of the non-thermal emission in the regions. We examine the spectral index trend towards the northern and southern regions of the halo using the $46\arcsec$-resolution spectral index image. The spectral indices are extracted from the northern and southern sectors shown in the \textit{left} panel of Figure~\ref{fig:spx_profile_halo}. The width of the radial sub-regions is taken to be approximately the size of the synthesised beam (i.e. $48\arcsec$ and $53\arcsec$ for the northern and southern sectors, respectively). The corresponding spectral index profiles in the \textit{right} panel of Figure~\ref{fig:spx_profile_halo} show that the spectral index does not change (i.e. within $1\sigma$) towards the directions of the X-ray SB jumps. This might be because the spectral index map with the resolution of $46\arcsec$ (140~kpc) cannot resolve the spatial spectral structure of the diffuse radio emission in this region of the halo. Although there are hints of a spectral flattening towards the regions of the X-ray SB jumps, high uncertainties in the spectral index measurements prevent us from drawing a conclusion from the trend. Future deep, high-resolution radio observations will be necessary. 
	
	\subsubsection{Radio and X-ray SB correlation}
	\label{sec:res_corr}
	
	\begin{figure*}
		\centering
		\includegraphics[width=0.325\textwidth]{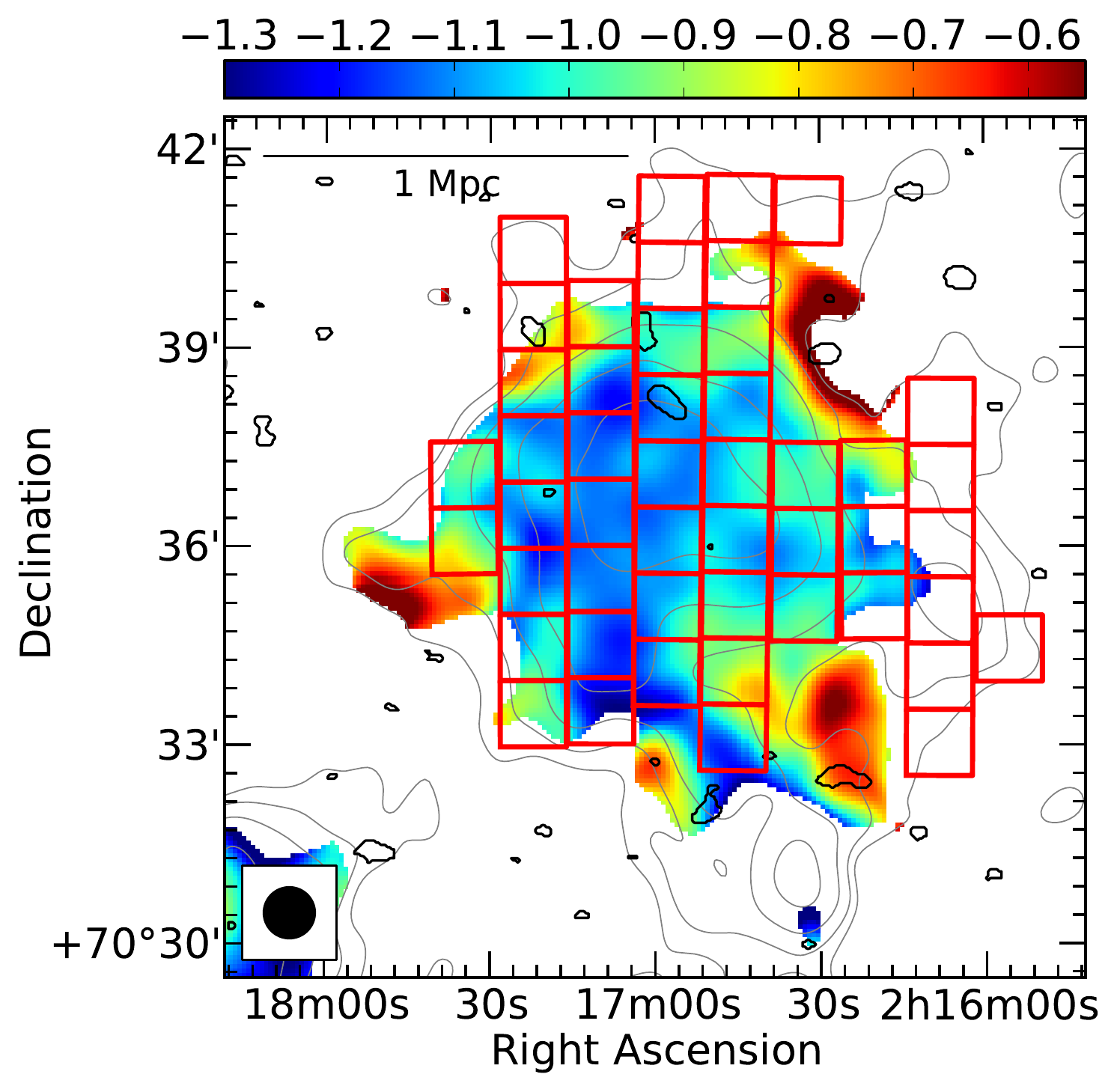} \hfill
		\includegraphics[width=0.325\textwidth]{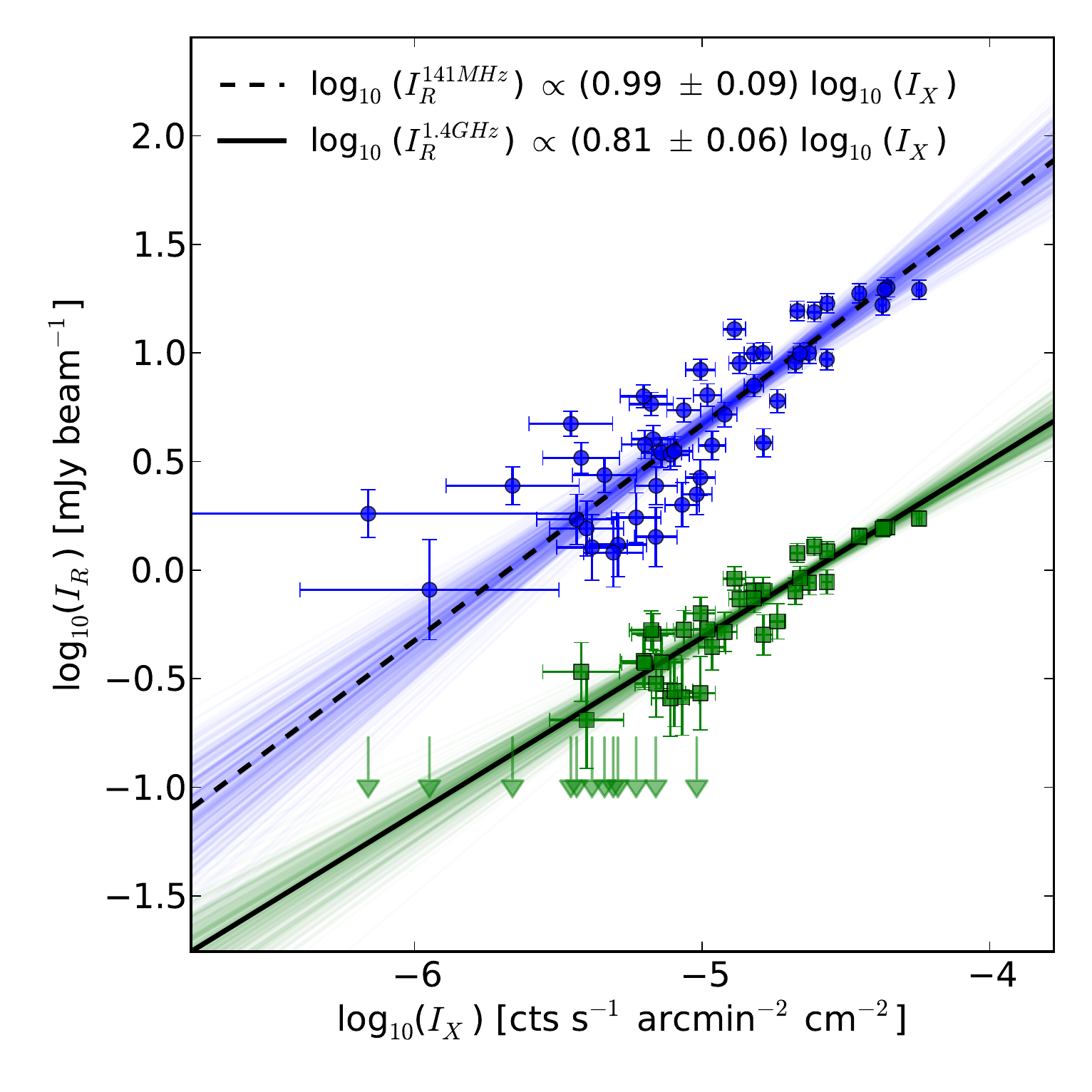}  \hfill
		\includegraphics[width=0.325\textwidth]{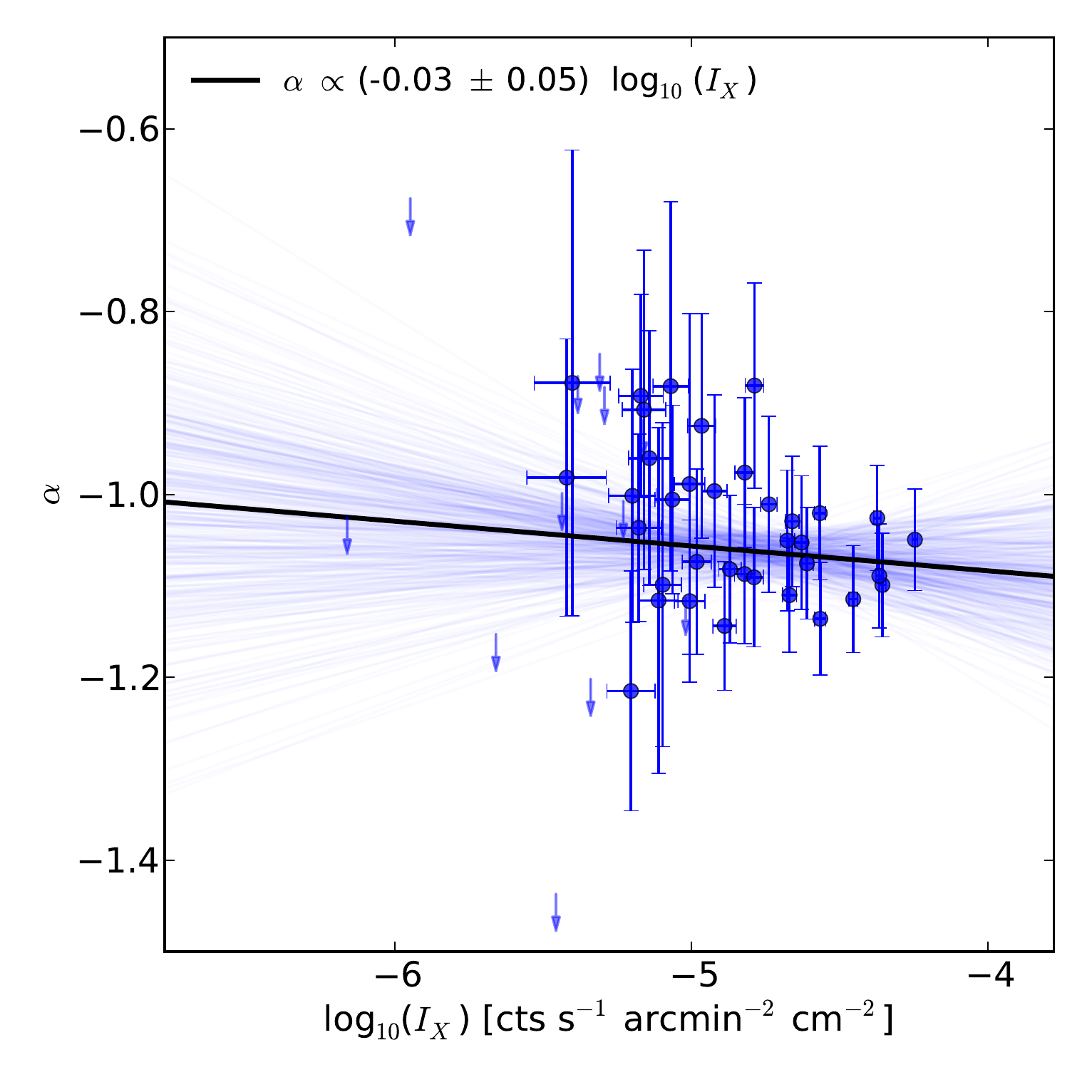} 	
		\caption{\textit{Left}: Regions where radio and X-ray data are extracted. \textit{Middle}: Correlation between radio and X-ray SB. \textit{Right}: Scatter plot of the radio spectral index and X-ray SB. The solid and dashed (black) lines show the best-fit results from the MCXC runs. The thin (blue and green) lines are the fitting results of some individual MCXC runs. In both plots, the $2\sigma$ upper limits are shown with the downward/leftward arrows. 
		}
		\label{fig:correlation}
	\end{figure*}
	
	%  correlation with X-ray emission
	With the current data, the X-ray emission from the ICM of ClG0217 is approximately as extended as the size of the radio emission, as seen in Figure~\ref{fig:df}. It is noted that the \textit{Chandra} data was obtained from the shallow 25~ks observations that might not detect the faint emission in the outer region of the cluster. The intensity of the X-ray and radio diffuse emission in the halo region is spatially correlated. This is clearest in the central region of the halo where elongated sub-structures in the X-ray and radio emission are seen in the NE--SW direction.
	
	The correlation between radio and X-ray emission in the cluster central region has been found to follow a power-law relation,
	\begin{equation}
		\log_{10}{I_R}= a + b\times \log_{10}{I_X},
	\end{equation}
	where $a$ and $b$ are free parameters \citep[e.g.][]{Govoni2001a,Feretti2001, Govoni2001c, Bruno2021, Rajpurohit2021}. The parameter $b$ describes how the X-ray and radio SB are correlated. For $b=1$, the SB of the thermal and non-thermal emission is linearly related over the source. Whereas, the SB values are sub-linearly and super-linearly related in cases of $b<1$ and $b>1$, respectively. In case of sub- or super-linear correlation, the non-thermal components radially decline slower or faster than the thermal ones. 
	
	We quantitatively examine the correlation between the thermal X-ray and non-thermal radio emission from the ClG0217 halo by performing a point-to-point analysis using the X-ray \textit{Chandra} and  radio images (i.e. LOFAR 141~MHz and VLA 1.4~GHz maps at $46\arcsec$ resolution). We extract the X-ray and radio surface brightness from 48 square regions that have sizes of $60\arcsec\times60\arcsec$ (i.e. $182\times182$~kpc$^2$), covering up to 1.5~Mpc$^2$ area of the halo. The grid used for extracting the X-ray and radio SB values is shown in the \textit{left} panel of Figure~\ref{fig:correlation}. The size of the each square region is set to be larger than the size of the synthesised beam which is $46\arcsec$. We note that the regions are selected based on the LOFAR $2\sigma$ contours but exclude the flat spectral index regions that are likely to be affected by the imperfect subtraction of compact sources. The \textit{Chandra} image is smoothed with a 2-dimensional Gaussian function that has a kernel of 5 pixels (i.e. $5\times1.968\arcsec$). 
	
	In the \textit{middle} panel of Figure~\ref{fig:correlation}, we show the point-to-point relation between the radio and X-ray SB. To check if there exists any significant correlation between these two quantities, we fit the observed data with a Bayesian linear regression using the $\mathtt{linmix}$ package\footnote{\url{https://github.com/jmeyers314/linmix}}. The $\mathtt{linmix}$ takes into account the uncertainties of both X-ray and radio data  \citep{Kelly2007} and also non-detections. For the halo in ClG0217, we find a  linear and \mbox{sub-linear} positive correlations at 141\,MHz and 1.4\,GHz, respectively. The correlation slope is $b_{\rm 141~MHz}=1.03\pm0.09$ and $b_{\rm 1.4~GHz}=0.81\pm0.06$. In Table \ref{tab:coeff}, we report the Pearson's (linear) correlation coefficients ($r$), including degree of freedom (DoF) and the $p$-value. The results suggest tight correlations between the radio and X-ray emission for the halo at both 141\,MHz and 1.4\,GHz. The obtained correlation slopes imply that the non-thermal components at 1.4~GHz declines more slowly than that at 141~MHz which is in line with the tentative flattened trend in our spectral index radial profile within the radius of 600~kpc, as discussed in Section~\ref{sec:res_spx}. % (see Figure~\ref{fig:spx_profile_halo}).
	
	\begin{table}
		\centering
		\caption{Parameters for the radio and X-ray SB correlation.
		} 
		\begin{tabularx}{\columnwidth}{@{}Xcccc@{}}
			\hline\hline
			& $b$            & $r$     & DoF   & $p$  \\ \hline
			LOFAR    & $1.03\pm0.09$  & 0.91    & 43     & <$0.001$ \\
			VLA      & $0.81\pm0.06$  & 0.97    & 34     & <$0.001$ \\
			$\alpha$ & $-0.02\pm0.06$ & $-0.21$ & 34 & $0.20$ \\
			\hline
		\end{tabularx}
		\label{tab:coeff} 
	\end{table}
	
	In a similar manner, we extracted the spectral indices in the square regions to study its relation with the X-ray SB that is fit with a relation of the form,
	\begin{equation}
		\alpha = a + b \times \log_{10}{I_X}.
	\end{equation}
	The resulting plot between the spectral indices and X-ray SB is shown in the \textit{right} panel of Figure~\ref{fig:correlation}. We obtained a slope of $b=-0.02\pm0.06$ and a Pearson's coefficient of $r=-0.21$, suggesting that the spectral index in the halo region is independent from the X-ray SB change. %This is consistent with the Pearson's correlation coefficient of $r=-0.21$ in Table \ref{tab:coeff}. 
	As the square regions in which the spectral indices are calculated are within a radius of 600~kpc, this result is in line with the spectral index profile shown in Figure~\ref{fig:spx_profile_halo}. We estimate the mean spectral index over the square regions in Figure \ref{fig:correlation} (\textit{left}) to be $-1.03\pm0.13$ which is consistent within $1\sigma$ with the integrated spectral index for the halo (i.e. $\alpha=-1.07\pm0.05$).

	\subsection{Peripheral diffuse sources}
	\label{sec:res_relics}
	
	\begin{figure*}
		\centering
		\includegraphics[width=0.393\textwidth]{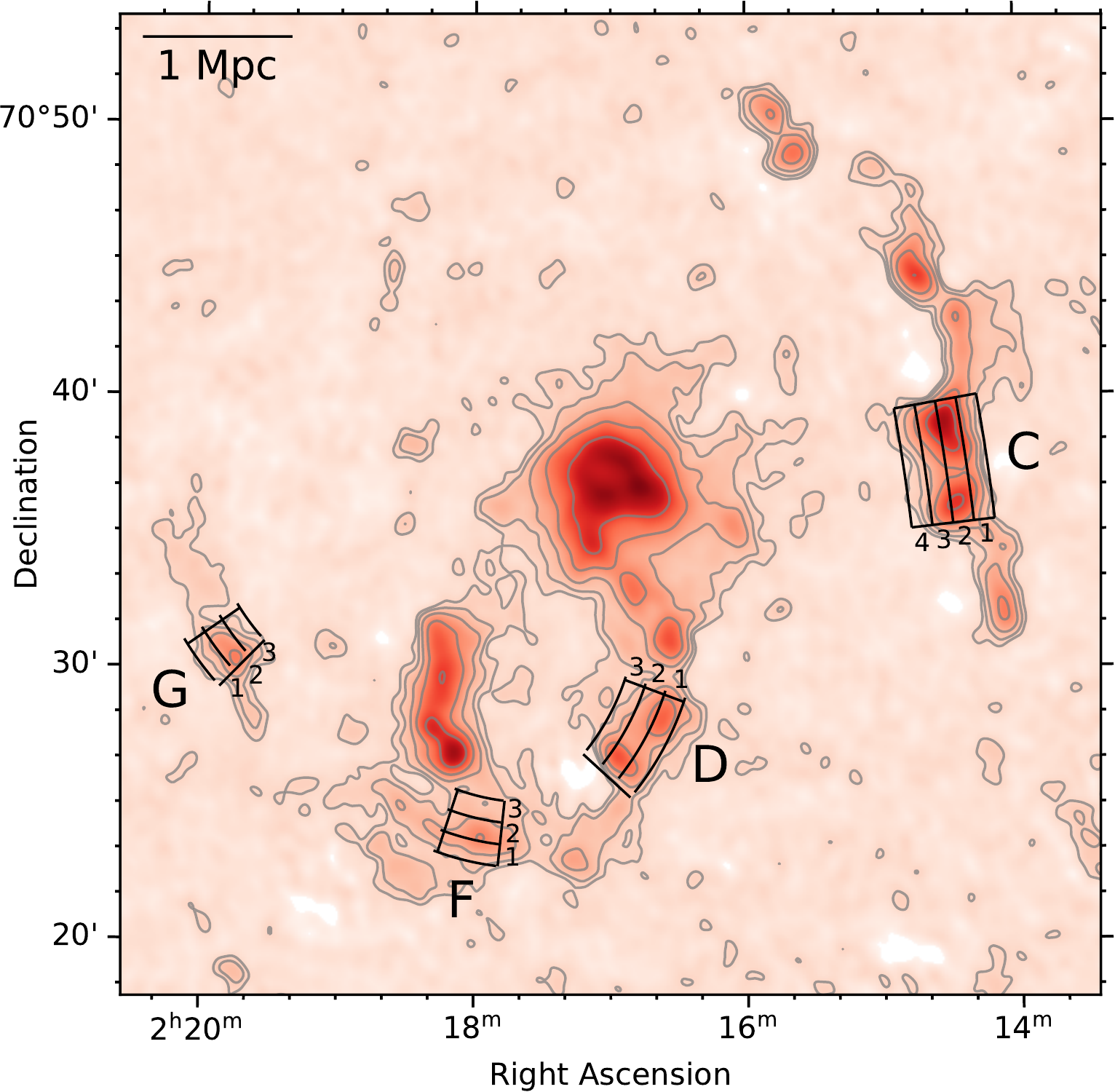} \hfill
		\includegraphics[width=0.58\textwidth]{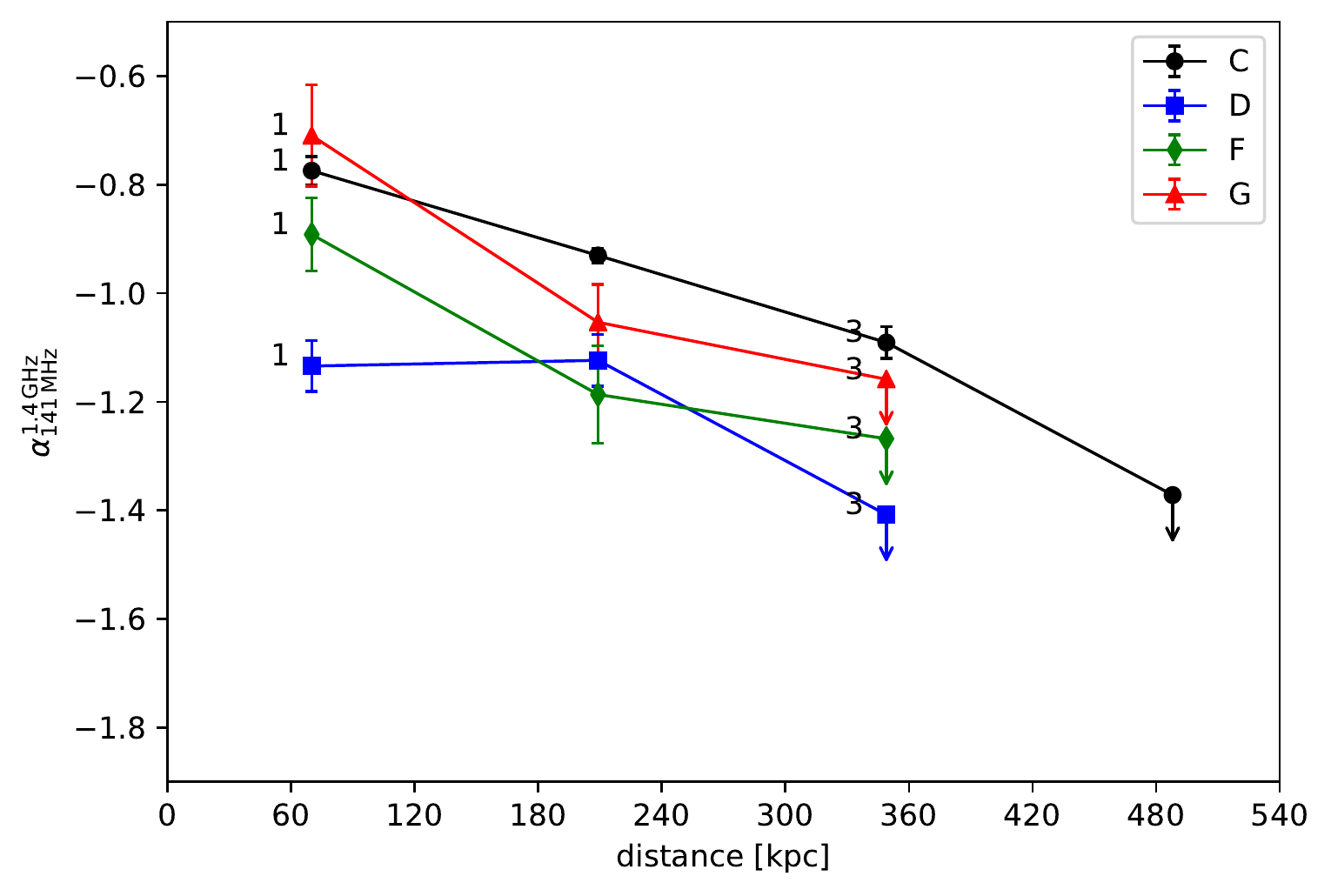}
		\caption{\textit{Left}: Regions across the relics where spectral indices are extracted. The width of the regions is $46\arcsec$ that is equal to the size of the synthesised beam. \textit{Right}: Spectral index profiles across the width of the relics on the \textit{left} panel. The downward arrows indicate the upper limit of the spectral indices. The error bars show the $1\sigma$ uncertainties that are estimated from the image noise. The upper limits for the spectral indices (i.e. the arrows) are calculated using the upper limits of $2\sigma$ for the 1.4~GHz emission.
		}
		\label{fig:spx_profile_relics}
	\end{figure*}
	
	The cluster ClG0217 is known to host several diffuse sources in the outskirts. Source C, for labelling see Figure~\ref{fig:label}, located at $15\arcmin$ to the west of the cluster centre, was classified as a radio relic by \cite{Brown2011a}. The measured projected length of the relic is about 1.7~Mpc at 1.4~GHz. In our deep LOFAR images, the relic C is 35 percent longer with a projected size of 2.3~Mpc. For these measurements, we do not include the bright, compact source to the north of the relic. A Panoramic Survey Telescope and Rapid Response System \citep[Pan-STARRS; ][]{Flewelling2020} optical counterpart in the radio emission peak suggests that the radio emission to the north of the relic is originated from a radio galaxy. We present a Pan-STARRS image of the optical source overlaid with LOFAR contours to show the connection in Appendix~\ref{sec:app_optical}. At 141\,MHz, the width of the relic C in projection is about 620~kpc in the middle region, as shown in Figure~\ref{fig:df}, and it gets narrower (i.e. 280~kpc) towards the south and north directions. In the north-western region of the relic, a diffuse, faint source, namely C1, is detected in the LOFAR low-resolution ($45.7\arcsec\times44.7\arcsec$) image, but it is not seen in the VLA images, that are shown in Figure~\ref{fig:vla_images}. We estimate that the 141~MHz--1.4~GHz spectral index of C1 is steeper than $-1.7$. 
	
	For the relic C, we measure a flux density at 141~MHz and 1.4~GHz within the LOFAR $2\sigma$ region and summarise these measurements in Table~\ref{tab:source_properties}. The region used for the flux density measurement is shown in Figure~\ref{fig:label}. Using these flux measurements, we estimate the integrated spectral index of the relic C to be $-1.01\pm0.05$ that is within typical range for known radio relics \citep[e.g.][]{VanWeeren2019a}. However, this value is significantly flatter than that of C1, which may imply their different origins. In Figure~\ref{fig:spx}, the relic C shows a clear spectral steepening in the downstream regions, towards the cluster centre. In the middle region, the spectral index decreases from $-0.72$ on the western side to $-1.35$ on the eastern side. We also extract the spectral index profile across the relic C and present it in the \textit{right} panel of Figure~\ref{fig:spx_profile_relics}. The spectral indices are estimated within the $46\arcsec$-width regions using the LOFAR 141~MHz and VLA 1.4~GHz $46\arcsec$-resolution images.These regions are shown in the \textit{left} panel of Figure~\ref{fig:spx_profile_relics}. The spectral steepening is more visible in the spectral index profile extracted across the relic C. This result supports the classification of the source as a radio relic.
	
	Other outskirts diffuse sources D, F, and G, for labelling see Figure~\ref{fig:label}, are more extended at 141~MHz compared to their sizes at 1.4~GHz. We refer to Table \ref{tab:source_properties} for the measurements of their extension. Due to its elongated shape and location, source D was thought to be an internal filament \citep{Brown2011a}. In our LOFAR image, e.g., in Figure~\ref{fig:spx_profile_relics} (\textit{left}), the source further extends in the NW--SE direction and its north-western part is connected with the radio halo while its south-eastern region is connected with source F. The extension of source D along its width (i.e. 600~kpc in the NW--SE direction) suggests that it does not entirely consist of discrete sources. 
	
	As seen in Figure~\ref{fig:df}, the sources F and G are oriented in the east-west and north-south directions, respectively. On the basis of VLA 1.4\,GHz images, \cite{Brown2011a} identified both of these sources as radio relics. In the LOFAR image, source D is connected with source F. The clear connection between D and F suggests that they belong to a single structure with a projected size of 3.5~Mpc. If source G is also included, this increases to 5.1~Mpc. On the outer region of source F towards the south-east, an excess of diffuse emission, namely F1, is detected only at 141~MHz. We find that the spectral index of F1 is steeper than $-1.7$. The flux densities and spectral indices for these sources are reported in Table~\ref{tab:source_properties}. The spectral index map presented in Figure~\ref{fig:spx} reveals a hint of the spectral steepening across the sources D, F, and G that is also seen in the spectral index profiles in Figure~\ref{fig:spx_profile_relics} (\textit{right}). The spectral indices across the relics are calculated in the similar manner as done for the source C (i.e. using the LOFAR~141~MHz and VLA~1.4~GHz $46\arcsec$-resolution images and $46\arcsec$-width regions). Based on the spectral and morphological properties of source D, we suggest that that it is a radio relic. Our results on the spectral steepening of F and G also support the relic classification of these sources by \cite{Brown2011a}.
	
	Due to the low-resolution of the VLA 1.4~GHz, sources B1+B2 (namely B in \citealt{Brown2011a}) were classified as a single radio relic by \cite{Brown2011a}, but in our high-resolution image in Figure~\ref{fig:lofar_hres}, it is composed of a bent-tail radio galaxy to the north (i.e. B1) and a radio galaxy with two lobes to the south (i.e. B2). These radio galaxies are also detected with the SDSS and Pan-STARRS optical surveys. The Pan-STARRS cut-out image in the B1+B2 region is shown in Appendix~\ref{sec:app_optical}. The SDSS identification is J021558.09+705040.1 ($z=0.1962\pm0.0802$) and J021542.94+704832.3 for the northern and southern sources, respectively.
	
	Source E1+E2 was unresolved in the VLA~1.4~GHz image and was identified as a single internal filament, namely E, by \cite{Brown2011a}. In our high-resolution LOFAR 141~MHz and VLA~1.5~GHz images shown in Figures \ref{fig:lofar_hres} and \ref{fig:vla_images} (\textit{left}), the source is resolved showing two parts (namely, E1 and E2). The southern part (E1) has a morphology typical of a double-lobe radio galaxy that has an SDSS optical counterpart (namely, J021814.33+702724.0 at R.A.=$02^{\rm h}18^{\rm m}14.33^{\rm s}$, Dec.=$+70^{\rm d}27^{\rm m}24.07^{\rm s}$) reported in the SDSS catalogue \citep{Alam2015}. In Appendix~\ref{sec:app_optical}, the cut-out Pan-STARRS colour image overlaid with the LOFAR contours shows better the optical counterpart for E1  \citep[with object identification: 192540345597208886;][]{Flewelling2020}. No redshift information is available for this optical source in the SDSS and Pan-STARRS catalogues. The morphology of the lobes suggests that it could be a typical Fanaroff-Riley Class I (FR-I) galaxy. The southern lobe of the active galactic nucleus (AGN) is bent towards the east. The northern part (E2) has an elongated shape that connects with a lobe of the southern radio galaxy. We interpret the source E being a radio galaxy that is moving in the NW--SE direction and leaving the tail of radio emission behind.
	
	%%%%%%%%%%%%%%%%%%%%%%%%%%%%%%%%%%%%%%%
	\section{Discussions}
	\label{sec:dis}
	
	\subsection{Radio halo}
	\label{sec:dis_halo}
	
	The presence of the radio halo in ClG0217 is confirmed with the LOFAR 141~MHz and VLA 1.5~GHz observations. The projected diameter of the halo at 141~MHz is 1.8~Mpc which is more extended than previously detected at 1.4~GHz \citep{Brown2011a}. 
	The integrated spectral index for the halo between 141~MHz and 1.4~GHz is $-1.02\pm0.05$.  %which is flatter than that estimated in \cite{Brown2011a}. 
	We discuss the spectral index properties and the correlation between thermal and non-thermal components below.
	
	\subsubsection{Spectral index variations}
	\label{sec:dis_halo_spx}
	
	% overall picture
	Spatial distribution of the spectral index in halos provides insight into the physical mechanisms responsible for the particle acceleration and magnetic field amplification in radio halos. In general, the spatial distribution of the spectral index provides information on turbulent scales, transport of particles and the magnetic field in the ICM.
	For example, homogeneous models where turbulence and magnetic field strength are uniform in the emitting volume predict a uniform distribution of the spectral index. On the other hand, models with inhomogeneous turbulence and magnetic fields predict variations in the spectral indices in the entire emitting volume. The variations depend on how fast transport of particles, diffusion of magnetic field, and turbulent scales are.
	
	% what are known?
	Observations of radio halos have shown conflicting radial distributions of the spectral index. A few cases have been found where the average radial spectral index steepens in the peripheral regions (i.e. Abell~2744; \citealt{Pearce2017}, MACS~J0717.5+3745; \citealt{Rajpurohit2021}). In some other cases, no firm detection of radial spectral steepening feature is detected (e.g. Coma~C\footnote{\label{note}The uncertainty of the spectral measurement in these cases is high due to the mismatching of the $uv$-coverages in the observations, as pointed out by \citealt{Botteon2020a}.}; \citealt{Giovannini1993}, Abell~665\footnoteref{note}, Abell~2163\footnoteref{note}; \citealt{Feretti2004b}, Abell~2219; \citealt{Orru2007}). In some ongoing merging clusters, a uniform distribution of the spectral index over a large fraction of the halos was also detected  (e.g. the Toothbrush cluster; \citealt{VanWeeren2016b,Rajpurohit2018,Gasperin2020,Rajpurohit2020}, the Sausage cluster; \citealt{Hoang2017,Gennaro2018}, and Abell~520; \citealt{Hoang2019a}). %, MACS J0717.5+3745; \citealt{Rajpurohit2021}).
	
	% ClG0217?
	In ClG0217, a spectral steepening is detected beyond 830~kpc from the cluster centre, as shown in Figure~\ref{fig:spx_profile_halo} (\textit{right}). However, within $680\,{\rm kpc}$, the radial profile is roughly constant although there is a hint of radial spectral flattening. We note that X-ray SB discontinuities are detected at the edges of the northern and southern regions \citep[within $\sim$600~kpc,][]{Zhang2020}. Although the nature of the X-ray discontinuities (i.e. shock or cold front) is still unknown, they might affect the observed radio spectrum of the halo through the \mbox{(re-)acceleration} of the CRs and/or amplification of magnetic field. In some clusters, shock fronts are observed at the edges of radio halos, including Abell~520 \citep{Markevitch2005}, the Bullet cluster \citep{Shimwell2014}, the Coma cluster \citep[i.e. the western edges of the halo; ][]{Brown2011b}, and the Toothbrush \citep[i.e. the southern edge; ][]{VanWeeren2016b}. Among these cases spectral flattening at the edges has been seen in two clusters, including the south-western region of Abell~520 \citep{Hoang2019a} and the southern shock of the Toothbrush \citep{Rajpurohit2018}. The halo in ClG0217 might be the third case if future deep radio observations confirm the radial spectral trend. 
	
	\subsubsection{Thermal and non-thermal correlation}
	\label{sec:dis_corr}
	
	The radio and X-ray emission in the halo of ClG0217 are found to be spatially correlated. In particular, as shown in Figure~\ref{fig:df}, the  radio and X-ray sub-structures in the central region of the halo are found to have similar shapes and orientations. Quantitatively, the radio and X-ray SB is consistent with the linear relation at 141~MHz with a correlation slope of $b_{\rm 141\,MHz}=1.03\pm0.09$. This slope is found to be \mbox{sub-linear} at 1.4~GHz with $b_{\rm 1.4\,GHz}=0.81\pm0.06$. The spatial correlation suggests a tight connection between the thermal gas and non-thermal components (i.e. relativistic electrons and magnetic field) in the ICM. Our results are similar to some cases where the radio and X-ray relation is linear/sub-linear (i.e. $b\sim0.64-1$) regardless of the morphology of the radio halos \citep[e.g.][]{Feretti2001,Govoni2001a, Govoni2001c, Bruno2021, Rajpurohit2021,Botteon2020a}. Also, a lower value of $b$ at higher frequency suggests additional non-thermal processes in the outer regions of the halo where X-ray emission is fainter. 
	
	In the halo of ClG0217, the \mbox{sub-linear} slope at high frequencies could be related to the presence of  X-ray SB discontinuities at the edges of the halo that might lead to a flattening of the spectrum (e.g. via adiabatic compression) and thus generate more flux in the periphery at 1.4~GHz. The hint of this trend (i.e. lower $b$ at higher frequencies) is also seen in Abell~520 which hosts two X-ray SB jumps (i.e. a strong bow shock and a tentative shock/cool front) at the halo edges in \cite{Hoang2019b}. They found $b_{\rm 145\,MHz}=0.34\pm0.11$, $b_{\rm 323\,MHz}=0.27\pm0.10$, and $b_{\rm 1.5\,GHz}=0.25\pm0.09$. However, an opposite trend (i.e. $b_{\rm 141\,MHz}=0.67\pm0.05$, $b_{\rm 1.5\,GHz}=0.81\pm0.09$, and $b_{\rm 3\,GHz}=0.98\pm0.09$ ) is also observed in the halo of MACS~J0717.5+3745 in  \cite{Rajpurohit2021}. The difference is likely due to the difference in the physical conditions in the two halos. % repetitive
	
	The simplest approach to turbulent \mbox{re-acceleration} models is to assume homogeneous conditions in the ICM. In this case, a broad spatial distribution of relativistic seed electrons and of the turbulence are assumed to be uniformed over a Mpc$^3$ volume. The synchrotron power is,
	\begin{equation}
		P \propto \rho \frac{\delta V^3}{L} \frac{B^2}{B^2 + B_{\rm CMB}^2}, 
	\end{equation}
	where $\rho$ is is the thermal particle density, $\delta V$  is the turbulent velocity, and $L$ is the corresponding turbulent scale \citep[e.g.][]{Brunetti2020a}. In the dynamo conditions, $B^2 \propto \rho dV^2$, and for $B<B_{\rm CMB}=3.2\times(1+z)^2\,{\rm \upmu G}$ the synchrotron power scales with $P \propto \rho^2$ implying a linear radio--X-ray correlation \citep[][]{Beresnyak2012,Brunetti2020a}. As a consequence, the linear correlation that is observed at 141~MHz between X-ray and synchrotron SB would imply a magnetic field smaller than $4.5\,{\rm \upmu G}$ in the halo volume; stronger fields would indeed generate a \mbox{sub-linear} scaling correlation. In this case, the \mbox{sub-linear} coefficient that is measured at 1.4~GHz may imply that additional mechanisms (e.g. compression at the edges of the halo) contribute to the \mbox{re-acceleration} of particles in the outer regions of the halo (i.e. low X-ray SB regions). This would also be consistent with the hint of a radial spectral flattening discussed in Section \ref{sec:res_spx}.

	% linear relation in: A2255, A2744
	% \mbox{sub-linear} relation in:
	% Feretti2001 A2136 (b =0.64 ± 0.05)
	% Govoni2001a A2255 (b =0.98 ± 0.04.), Coma (b =0.64 ± 0.07), A2744: b =0.99 ± 0.05,  A2319 b =0.82 ± 0.04
	% Govoni2001c (no spatial correlation) 
	% Shimwell2014 RX positive correlationi, alpha slope: negative correlation
	% Rajpurohit2018 RX: b = 1.25 ± 0.16
	% Hoang2019 (b = 0.34 ± 0.11, 0.27 ± 0.10, and 0.25 ± 0.09 at 145 MHz, 323 MHz, and 1.5~GHz), alpha no slope, alpha = -1.01
	% Cova2019: RX: 1.156 ± 0.053 (A2744), 0.095 ± 0.08 (A523) 
	% Xie2020 RX: 150 MHz 0.55±0.04, 1.5~GHz 0.48 ± 0.07 
	% Rajpurohit2020 (MACSJ0717.5+3745:
	%144 MHz, 1.5~GHz and 3GHz are 0.67 ± 0.05, 0.81 ± 0.09, and 0.98 ± 0.09 ; slope alpha -0.09 and -0.20 (negative)
	% Botteon2020a (slope RX: 1.06 ± 0.15, slope alpha, 0.51 ± 0.10, alpha=-1.34) 
	% tight: Feretti2001, A2163, b =0.64 ± 0.05
	% sub-linear: 
	
	\subsection{Radio relics} 
	\label{sec:dis_relics}
	
	\subsubsection{Shock Mach numbers}
	\label{sec:dis_M}
	
	% shock model
	The elongated shape, location, and the lack of optical counterparts of C, D, F, and G suggest that they are radio relics. The spectral steepening across their width shown in Figure~\ref{fig:spx_profile_relics} further supports the classification. In the context of direct shock acceleration model, the relativistic electrons emitting the radio synchrotron emission in radio relics are accelerated directly from thermal pool at shock fronts. The energy distribution of these relativistic electrons follows the standard power-law function $\nicefrac{dN(p)}{dp} \propto p^{-\delta_{\rm inj}}$, where $dN(p)$ is the  number of particles in the energy range between $p$ and $p+dp$, and $\delta_{\rm inj}=1-2\alpha_{\rm inj}$ is the energy index. The spectral index of the injected relativistic electrons, $\alpha_{\rm inj}$, non-linearly depends on the Mach number of the shocks \citep[e.g.][]{Blandford1987}, 
	\begin{equation}\label{eq:M}
		\mathscr{M} = \sqrt{\frac{2\alpha_{\rm inj}-3}{2\alpha_{\rm inj} +1 }},
	\end{equation}
	where $\alpha_{\rm inj}>-0.5$. After being injected from the shock fronts the relativistic electrons start to lose their energy in the downstream region due to the synchrotron and inverse-Compton radiation, resulting steeper integrated spectral index over the source volume \citep{Ginzburg1969},
	\begin{equation}\label{eq:alpha}
		\alpha_{\rm int} = \alpha_{\rm inj} - 0.5.  
	\end{equation}
	
	\begin{table}
		\centering
		\caption{Shock Mach numbers for radio relics.
		}
		\begin{tabularx}{\columnwidth}{@{}Xcccc@{}}
			\hline\hline
			Source  & $\alpha_{\rm inj}$ &        $\mathscr{M_{\rm inj}}$  &   $\alpha_{\rm int}$ &  $\mathscr{M_{\rm int}}$   \\
			\hline
			C       &    $-0.72\pm0.05$     &   $3.2^{+0.4}_{-0.3}$   & $-1.01\pm0.05$ & 14 \\
			D       &      $-1.14\pm0.07$      &    $2.0^{+0.1}_{-0.1}$ & $-1.34\pm0.06$ & $2.6^{+0.2}_{-0.2}$\\
			F       &     $-0.93\pm0.08$    &     $2.4^{+0.2}_{-0.2}$ & $-1.09\pm0.06$ & $4.8^{+3.4}_{-1.0}$  \\
			G       &     $-0.97\pm0.16$    &      $2.3^{+0.4}_{-0.2}$ & $-1.00\pm0.06$ & $\gtrsim5.8$  \\ 
			\hline
		\end{tabularx}\\
		\label{tab:shock_properties}
	\end{table}
	
	% M from inj and int
	The Mach numbers can be estimated through the discontinuity of X-ray SB or temperature jumps that associate with the particle density across the shock fronts. However, the X-ray \textit{Chandra} data we use in this study is too shallow or does not cover the regions of the radio relics which does not allow us to search for the shocks at these regions with the X-ray data. To estimate the strength of the shocks associated with the relics, we calculate their Mach numbers using the injected spectra that are measured in the injection regions and the volume of the sources. In both methods, we use the LOFAR 141~MHz and VLA~1.4~GHz $46\arcsec$-resolution images (see Table \ref{tab:image_para} for image properties). We measure the injection spectral indices in the outer regions of the relics C, D, F, and G that have flattest spectra where the relativistic electrons are injected. Taking into account the image resolution, we set the width of the extracted regions to be the same as the beam size (i.e. $46\arcsec$).  The corresponding Mach numbers are estimated using Equation~\ref{eq:M} and given in Table \ref{tab:shock_properties}. The Mach numbers for these relics range between $2.0$ and $3.2$ that are typically found for shocks in merging clusters of galaxies \citep[e.g.][]{Markevitch2007}. 
	
	Alternatively, the Mach numbers can be calculated from the volume integrated spectral index via  Equations~\ref{eq:M} and \ref{eq:alpha}. We use the integrated spectral indices for the relics in Table \ref{tab:source_properties}. The resulting Mach numbers for D and F are $\mathscr{M}_{\rm D}^{\rm int}=2.6^{+0.2}_{-0.2}$ and  $\mathscr{M}_{\rm F}^{\rm int}=4.8^{+3.4}_{-1.0}$, respectively. These estimates are higher than those we estimated with the injection spectral indices, as summarised in Table \ref{tab:shock_properties}. Previous studies have also found this systematic offsets between the estimates that use spatially resolved maps and integrated spectra \citep[e.g.][]{Stroe2013a,Hoang2017,Hoang2018}.  However, the integrated spectral index method when used for the relics C and G results in unrealistic Mach numbers for merger shocks. For instance, the integrated spectral index for C is $-1.01\pm0.06$ that corresponds to a shock Mach number of $\mathscr{M}_{\rm C}^{\rm int}=14$ which cannot be found in merging galaxy clusters as it requires very different particle acceleration efficiency than those in other relics D and F. For the relic G, the integrated spectral index of $-1.00\pm0.06$ implies an injection spectral index of $-0.50\pm0.06$ which cannot be used for Equation~\ref{eq:alpha} (i.e. square root of a real negative number). This implies that the relic G is a clear example where Equation~\ref{eq:alpha} is invalid. In other words, the cooling time of the radio-emitting particles is longer than crossing time of the shock. In case of the lowest $1\sigma$ limit on the integrated spectral index (i.e. $\alpha=-1.06$), the lower limit for the Mach number is $\mathscr{M}_{\rm G}^{\rm int}\gtrsim5.8$.
	
	% pros and cons of both
	The estimate of the Mach numbers through the injection and integrated spectral indices has pros and cons, as discussed in Section 4.1.3 of \cite{Hoang2017}. In the former method, the injection index is mostly affected by the projection effect, beside the spatial resolution and the misalignment of radio images. The measurement bias for the injection index is reduced when the shock propagates on the plane of the sky as it minimises the mixture of relativistic electrons with different spectra in the downstream region. Although the plane of shock waves in ClG0217 is unconstrained, it is unlikely to deviate significantly from the plane of the sky as it is typically found for double-relic clusters. Hence, the injection spectral indices obtained directly from the spectral index map is likely to provide more reliable estimates of the shock Mach numbers associated with the relics. However, we still need further investigation with optical observations to constrain on the merger dynamics of the cluster. 
	
\begin{figure}
		\centering
		\includegraphics[width=1\columnwidth]{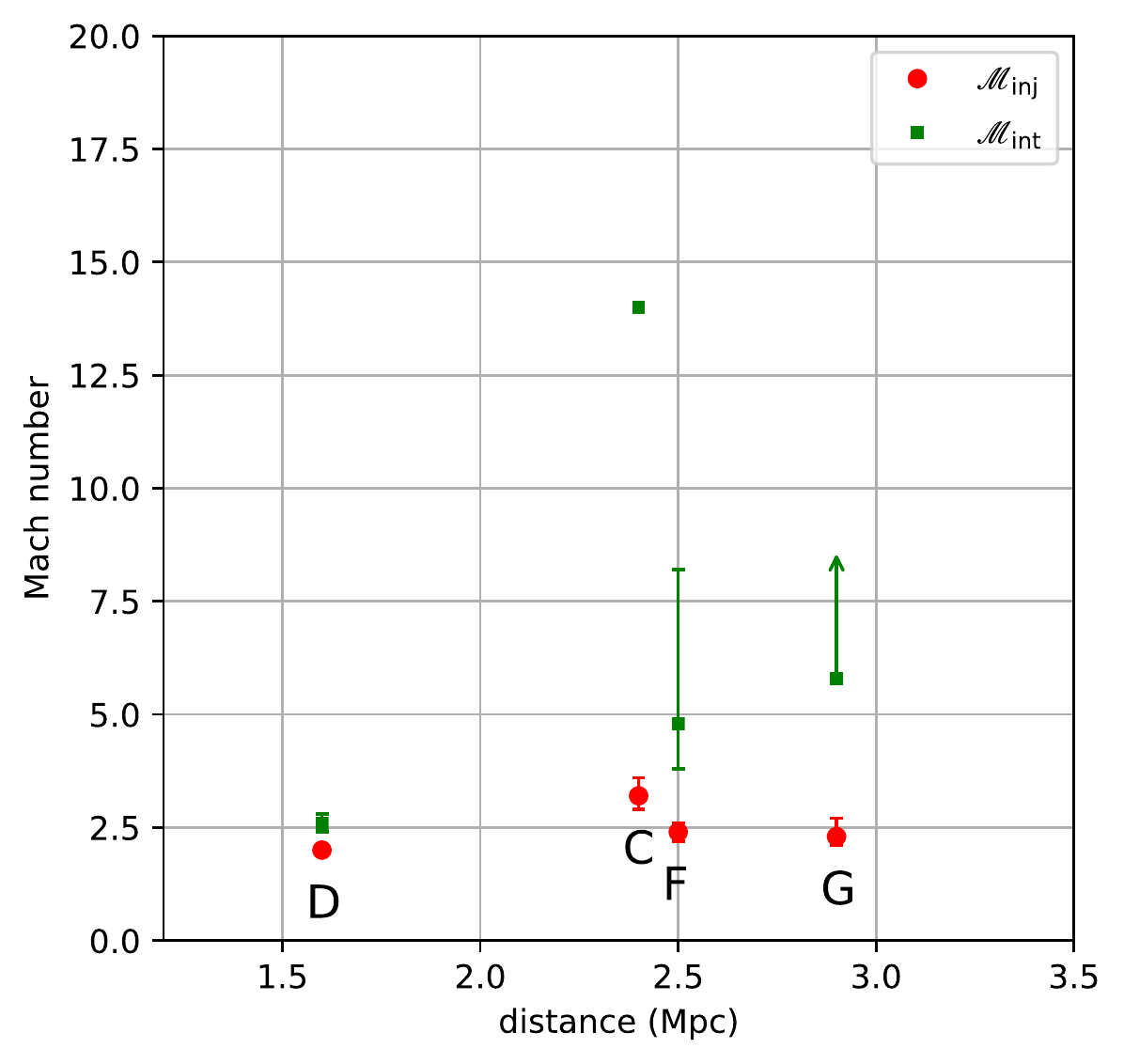}
		\caption{A scatter plot of the shock Mach numbers and the distance of the relics from the cluster centre. 
		}
		\label{fig:Mach}
	\end{figure}

	Moreover, the orientation of the major axes of the relics D, F, and G suggests that they are related to a structure which is supported by the Mach numbers of these relics not being largely different (i.e. $2.0-2.4$; Table \ref{tab:shock_properties}). When using the integrated spectral index, we are not biased by the spatial resolution, the projection effect, or the misalignment of the radio images. However, we assume that radio relics are caused by planar shock waves (i.e. Equation~\ref{eq:alpha}).  Simulations by \cite{Kang2015a,Kang2015b} found that this assumption does not hold for spherical shocks that are typically found in merging systems (due to their arc-like shape). The deviation significantly increases in time as the shock propagates towards the outskirts. This explanation is in line with the fact that the radial distance increases from $1.6\,{\rm Mpc}$ to $2.5\,{\rm Mpc}$ and $2.9\,{\rm Mpc}$ for D, F, and G, respectively. As a result, the Mach numbers inferred from the integrated spectral indices largely differ from those using the spectral index map (i.e. $\mathscr{M}_{\rm D}^{\rm int}=2.6^{+0.2}_{-0.2}$ and  $\mathscr{M}_{\rm F}^{\rm int}=4.8^{+3.4}_{-1.0}$, but it is unusable for the extreme case of the relic G). On the other hand, the Mach number discrepancy for C is more significant than that of F although the relic C is closer to the cluster centre than F. This could imply that the shocks in the west (for C) and south-east (for F) directions travels through different physical environments \citep{Dominguez-Fernandez2020}. In Figure~\ref{fig:Mach}, we show the Mach number obtained from integrated and injection index as a function of distance of these relics from the cluster centre. The discrepancy for the shock Mach numbers associated with the relics are clearly visible. 
	
	\subsubsection{Particle acceleration efficiency}
	\label{sec:eff}
	
	\begin{figure}
		\centering
		\includegraphics[width=1\columnwidth]{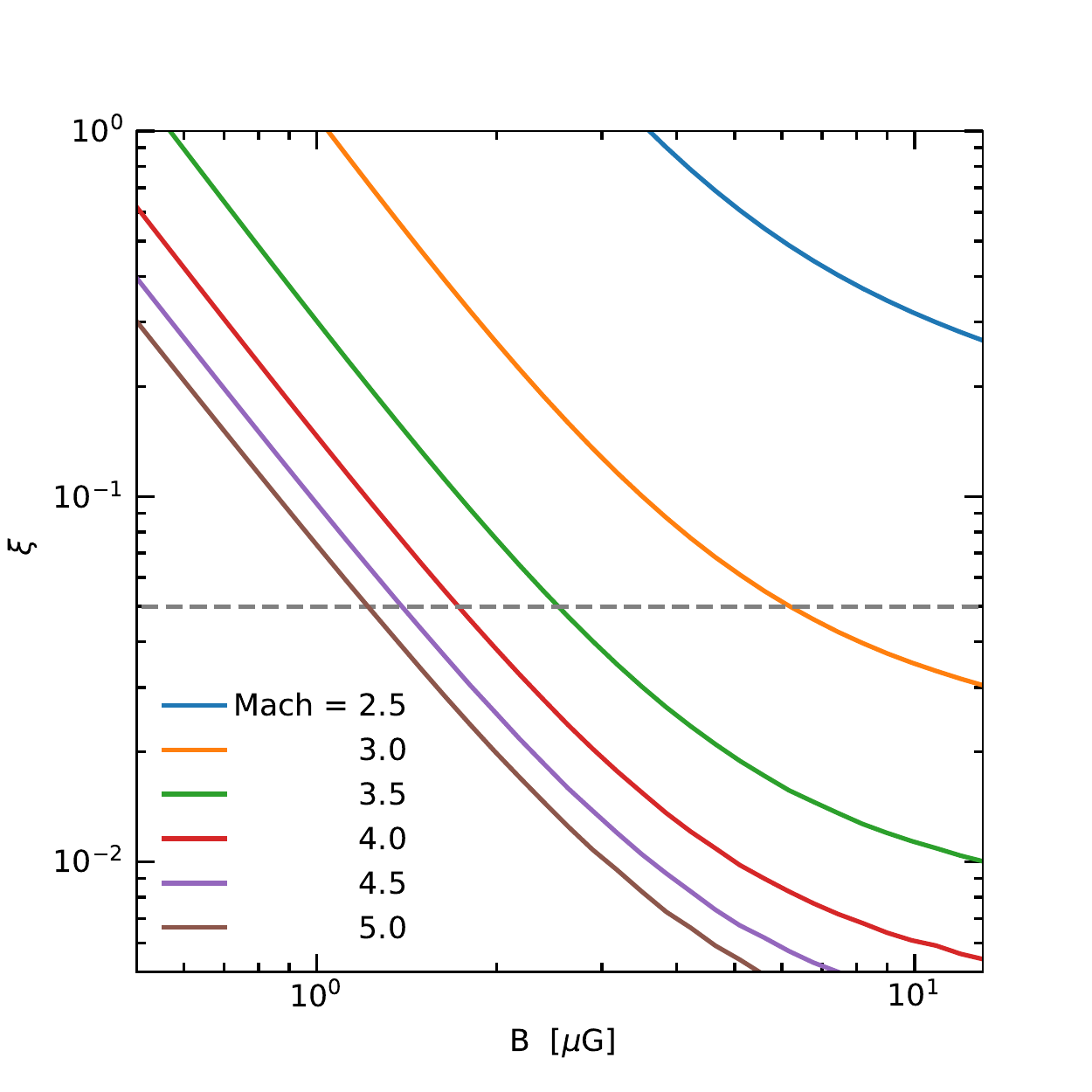}
		\caption{Particle acceleration efficiency $\xi$ as a function of the downstream magnetic field strength $B$ for the relic C with different Mach numbers. The horizontal dashed line indicates an efficiency of 5 percent.
		}
		\label{fig:eff}
	\end{figure}
	
	For relics associated with $\mathscr{M}\gtrsim2.5$ shocks, a direct acceleration of thermal electrons via DSA might be possible as it requires a lower acceleration efficiency \citep[$\xi\lesssim0.1$,][]{Botteon2020,Brunetti2014}. We explore the possibility of direct shock acceleration for the relic C that is related to the strongest  ($\mathscr{M}_{\rm C}^{\rm inj}=3.2^{+0.4}_{-0.3}$) shock in the cluster. In Figure~\ref{fig:eff}, we present the acceleration efficiency as a function of downstream magnetic field strength for the relic C with different Mach numbers, ranging between 2.5 and 5. The relation between the radio power of the relic, the efficiency, the Mach numbers and other properties of the downstream medium is based on Equation 32 of \cite{Hoeft2007a} taking into account a recent revision of the relation, for the updated  expression see \cite{Rajpurohit2021a}. In the calculation of the acceleration efficiency, we use an average temperature of 5~keV at the relic location (i.e. also $R_{200}$) that is estimated using a typical temperature profile for clusters \citep{Pratt2006} and a cluster average temperature of 8.3~keV \citep{Zhang2020}. The downstream electron density is estimated to be $2.6\times10^{-5}\,{\rm cm^{-3}}$, based on a typical gas density at the cluster virial radius.  The shock surface area is assumed to be the area of a spherical cone with an area of 4~Mpc$^2$. The radio power at 1.4~GHz is taken from the measurement of the VLA 1.4~GHz (i.e. $P_{1.4~{\rm GHz}}=1.85\times10^{24}\,{\rm W\,Hz^{-1}}$). With these parameters fixed the shock acceleration efficiency can be given as a function of downstream magnetic field and Mach number. 
	
	Figure~\ref{fig:eff} shows that the acceleration efficiency required to produce the relic brightness decreases when the downstream magnetic field and the shock Mach number increase. For instance, the efficiency needs to be less than 5 percent for Mach numbers of $\mathscr{M}\gtrsim3.5$ in the $B\gtrsim3$~$\upmu$G field. In these cases, a direct acceleration of thermal electrons might be still possible for the relic C although it depends on our assumptions on, e.g., the geometry of the shocks, the temperature profile of the cluster, and the typical gas density at the virial radius which we are unable to confirm with the present data. Nevertheless, a pre-existing population of mildly relativistic electrons prior to the re-acceleration of the shock will provide an additional source of energy for the relic lowering the required acceleration efficiency of the shock.
	
	\subsubsection{Shock re-acceleration of fossil electrons}
	\label{sec:dis_reacc}
	
	% For relics associated with low Mach number ($\mathscr{M}\lesssim2.5$) shocks, the efficiency of particle acceleration directly from the thermal pool is generally high  \citep[$\xi\gtrsim0.1$,][]{Botteon2020,Brunetti2014}. However, c
	Current theoretical models predict a smaller efficiency (i.e. a few percent) for low Mach number ($\mathscr{M}\lesssim2.5$) shocks \citep[e.g.][]{Kang2005,Kang2013}. A possibility for the generation of the relics by these shocks is that the radio-emitting relativistic electrons are \mbox{re-accelerated} from a population of mildly relativistic electrons from, e.g., previous cluster mergers, AGN activities, or supernovae \citep[e.g.][]{weeren2013,Bonafede2014,shimwell2015,Botteon2016a,VanWeeren2017}. In this re-acceleration scenario, the energy that is required to be transferred to the radio-emitting relativistic electrons decreases. Hence, the acceleration efficiency required to produce the brightness of the relics is lower. In addition, unlike the thermal population that smoothly distributes over the cluster volume, fossil plasma from, e.g., AGN activities typically has small-scale structure that is similar to that of the relics \citep[e.g.][]{Bonafede2014}.
	
	In ClG0217, the relics, excluding C, are associated with low Mach number ($\mathscr{M}=2.0-2.4$) shocks that are likely to be too weak for a direct acceleration of the thermal electrons and, as discussed, requires the presence of fossil plasma ahead of the shock. Supporting evidence for this scenario in ClG0217 is that the relics show a clumpy structure and extended radio emission in front of the shock fronts. Our result is in line with \cite{Brown2011a} where the re-acceleration of fossil electrons by AGN activities was pointed out to explain the variations in the radio SB of the relics (C, F, and G). In addition, the location of C1 and F1 in the LOFAR maps suggests a possible connection with the relics C and F. Furthermore, the steep ($\alpha<-1.7$) spectral properties of C1 and F1 support their nature being fossil plasma, rather than being caused by merger shocks. However, the present LOFAR observations do not detect diffuse faint emission in front of the entire length of the relics (C and F), suggesting that the radio emission of the fossil electrons in these regions is faint which requires deeper observations to confirm. 
	
	\subsection{Spectral variations over source E}
	\label{sec:dis_E}
	
	\begin{figure*}
		\centering
		\includegraphics[width=0.205\textwidth]{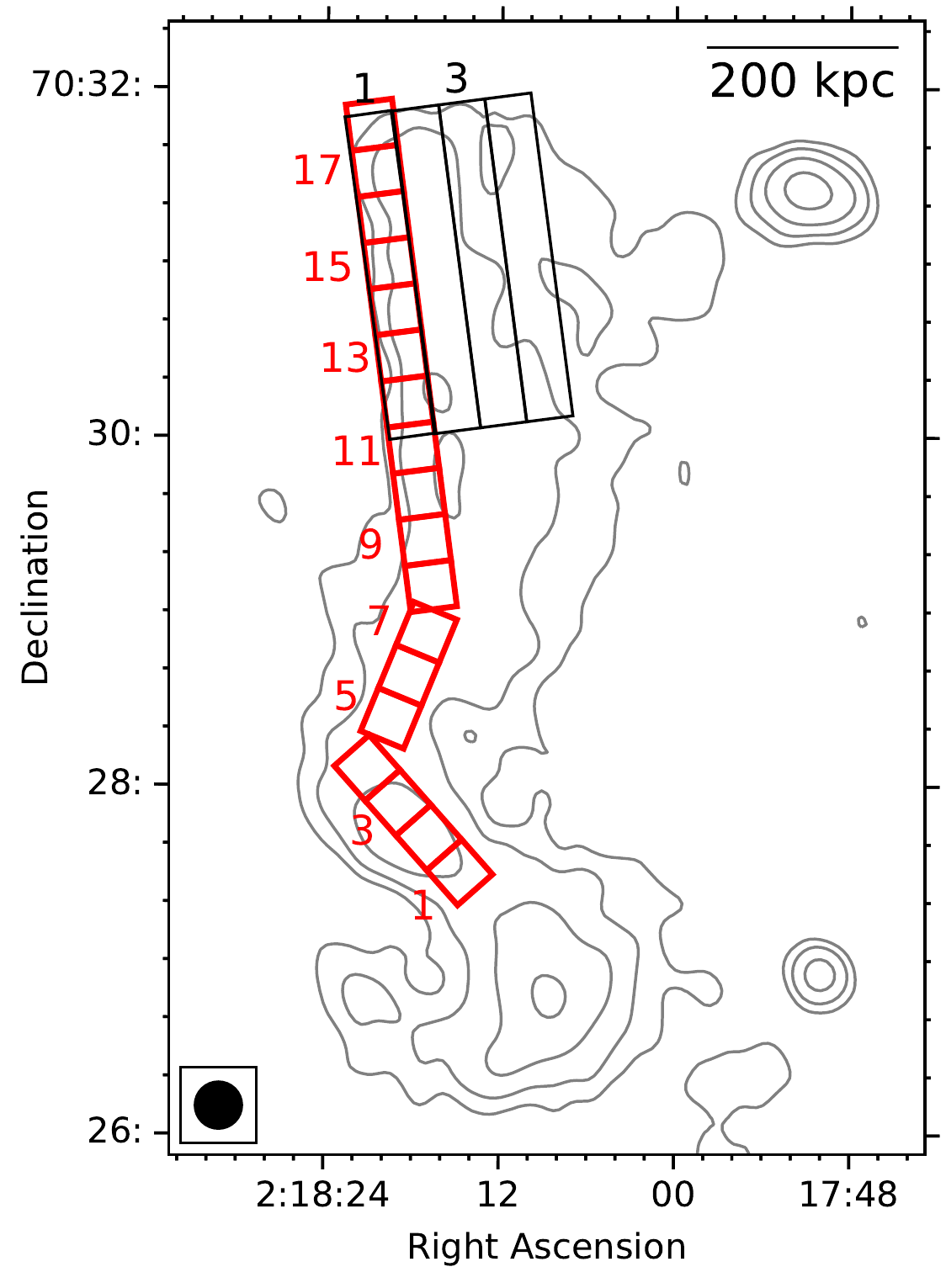} \hfill
		\includegraphics[width=0.385\textwidth]{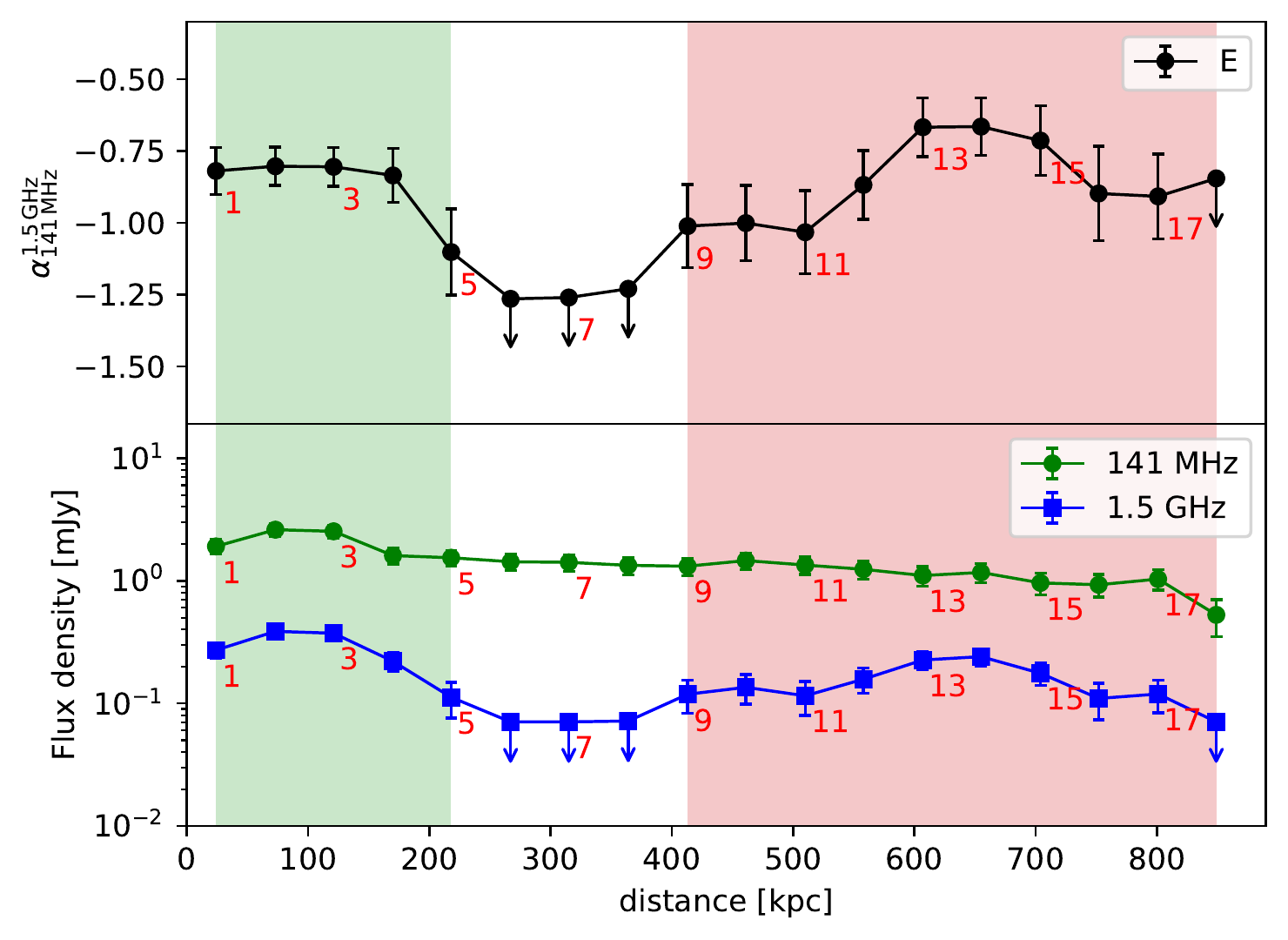}  \hfill
		\includegraphics[width=0.385\textwidth]{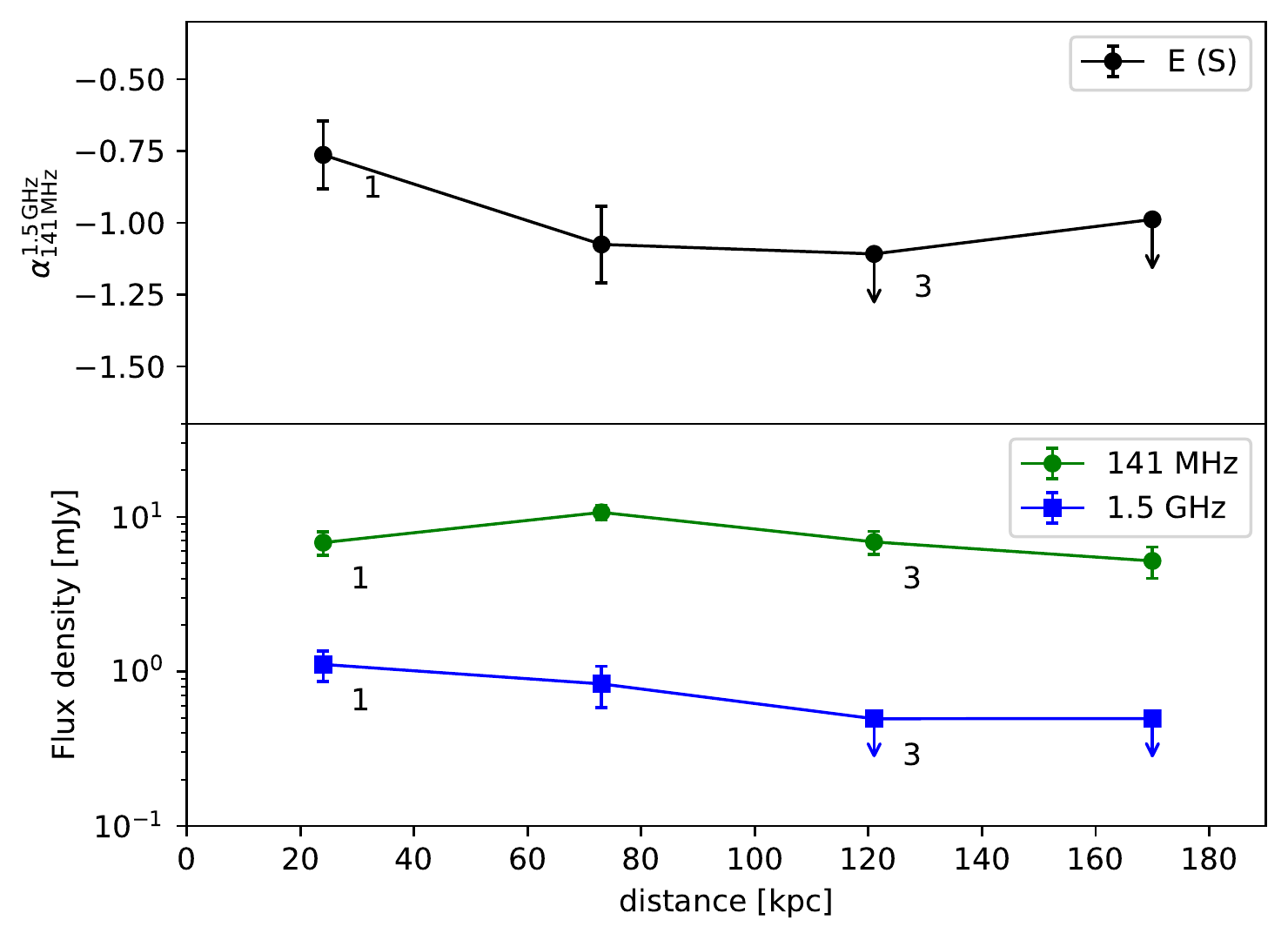} 	
		\caption{\textit{Left}: Regions over source E (i.e. E1+E2) where SB and spectral indices are extracted and are shown in the \textit{middle} and \textit{right} panels. \textit{Middle}: the SB and spectral index profiles along the SN direction (i.e. red boxes). \textit{Right}: Similar to those in the \textit{middle} panel, the profiles are calculated in the EW direction (i.e. black rectangles).
		}
		\label{fig:profile_E}
	\end{figure*}
	
	To examine the spectral index in source E, we extract 141~MHz--1.5~GHz spectra in the square regions from the centre of the southern part towards the north. These regions are shown in the \textit{left} panel of Figure~\ref{fig:profile_E}. The size of each region is equal to the beam size of the high-resolution spectral index, i.e. $16\arcsec\times16\arcsec$, corresponding a physical size of $49\,{\rm kpc}\times49\,{\rm kpc}$. The projected distance along these 18 regions is $870\,{\rm kpc}$. In the \textit{middle} panel of Figure~\ref{fig:profile_E},  we present the extracted spectra. The spectral index steepens from the central region of the AGN towards the northern lobe (i.e. from $-0.82\pm0.08$ in region 1 to $-1.10\pm0.15$ in region 5). In the regions between the southern and northern parts of E (regions $6-8$), the diffuse emission is undetected at 1.5~GHz, implying a steep spectral index below $-1.25$. The spectral steepening feature is typically observed in lobes of AGN which is due to the energy losses via synchrotron and inverse-Compton radiation. Further to the north, the spectral index becomes flat. The index increases up to $-0.68\pm0.11$ in the northern regions (i.e. regions $13-15$) where the source is significantly brighter at 1.5~GHz.
	
	The spectral flattening in the northern part of source E, i.e. E2, suggests the presence of a physical process that re-accelerates the fossil plasma of the southern AGN and/or amplifies the magnetic field in the regions. A possible origin of such process is caused by a merger shock that is propagating eastwards. One such example has been observed in the merging galaxy cluster Abell~3411-3412 \citep{VanWeeren2017}. In this case, fossil electrons from an AGN lobe lose their energy by spectral ageing before they are \mbox{re-accelerated} or adiabatically compressed by a merger shock. The re-acceleration leads to the spectral flattening of the radio emission in the region of the shock front. An additional feature of the shock re-acceleration is that the spectral index also steepens in the direction behind the shock front or towards the cluster centre as for outwards moving shocks. Some other examples for the relic-AGN connection and/or interaction between AGN in shocks include the southern relic of the Sausage \citep{Gennaro2018}, Abell~781 \citep{Botteon2019a}, the eastern relic of RXC~J1314.4--2515 \citep{Stuardi2019},  the northeastern relic in PLCK~G287.0+32.9 \citep{Bonafede2014}, and the southern relic of PSZ2~G096.88+24.18 \citep{Jones2021}.
	
	To examine the possible spectral ageing behind the northern part of E, we extract spectra in the regions across the width of its northern part (i.e. in east-west direction). These regions are shown in black in the \textit{left} panel of Figure~\ref{fig:profile_E}. The width of the each region is equal to the resolution of the LOFAR 141~MHz and VLA 1.5~GHz images, i.e. $16\arcsec$. The corresponding spectral index profile is shown in the \textit{right} panel of Figure~\ref{fig:profile_E}, suggesting a spectral steepening from east to west (i.e. from $-0.76\pm0.12$ to $-1.08\pm0.13$). This spectral steepening further supports our speculation that the northern part of source E is re-accelerated/compressed by an outwards moving shock. Using Equation~\ref{eq:M} and the spectral index of $-0.76\pm0.12$ in the eastern region of the northern part of E, we find that the Mach number of the possible shock is $\mathscr{M_{\rm E}}=2.9^{+1.0}_{-0.4}$. In this case, we use the integrated spectral index for the northern part (i.e. $-1.20\pm0.05$), the corresponding Mach number is $\mathscr{M_{\rm E}}^{\rm int}=3.3^{+0.5}_{-0.3}$ which is consistent within $1\sigma$ with the estimate directly using the spectral index map. Although the uncertainties of the Mach numbers are high, the consistency between the Mach numbers for the northern part of E might be due to the fact that the source E is closer to the cluster centre (e.g. compared with C, F, and G) where the approximation in Equation~\ref{eq:alpha} still holds. The presence of the shock front, if it exists, can be confirmed by future X-ray observations that we are unable to check with the current data.
	
	%%%%%%%%%%%%%%%%%%%%%%%%%%%%%%%%%%%%%%%
	\section{Conclusions}
	\label{sec:con}
	
	In this paper, we present LOFAR 141~MHz and VLA 1.5~GHz observations of the merging galaxy cluster ClG0217 that was previously studied with the VLA~325~MHz and 1.4~GHz by \cite{Brown2011a}. The LOFAR observations provide the first  radio images of the cluster at the frequencies below 325 MHz and confirm the presence of a giant radio halo and multiple radio relics in the cluster. To study the nature of these diffuse sources, we made spectral index maps of the cluster using the LOFAR 141~MHz and VLA 1.4~GHz/1.5~GHz data sets. We also made use of the archival \textit{Chandra} data to investigate the correlation between the thermal and non-thermal emission from the central radio halo and search for discontinuities in the X-ray SB. We summarise the results below:
	
	\begin{enumerate}
		
		\item The radio halo with a projected size of $1.8$~Mpc is observed to be more extended at 141~MHz than that at higher frequency of 1.4~GHz. Its flux density at 141~MHz is $623.6\pm62.7$~mJy, corresponding to a radio power of $P_{\rm 141\,MHz}=(56.9\pm5.7)\times10^{24}\,{\rm W\,Hz}^{-1}$.  The radio power of ClG0217 is roughly in line with the prediction from the radio power -- mass relation.
		
		\item The integrated spectrum of the radio halo in ClG0217 is $-0.78\pm0.16$ between 141~MHz and 325~MHz and steepens to $-1.16\pm0.07$ between 325~MHz and 1.4~GHz. This suggests the radio halo has a curved spectrum. However, further observations are needed to confirm due to the low resolution of the 325~MHz data which may lead to an imperfect subtraction of compact sources contaminating the halo.
		
		\item The non-thermal radio and thermal X-ray emission in the halo region where the radio and X-ray emission is detected above $2\sigma$ are positively correlated. At 141~MHz, the correlation is consistent with a linear relation, $\log_{10}{I_{R}} \propto b \times \log_{10}{I_{X}}$, where $b_{\rm\,141~MHz}=1.03\pm0.09$. However, at 1.4~GHz the correlation follows a \mbox{sub-linear} relation with $b_{\rm 1.4\,GHz}=0.81\pm0.06$. The results suggest that the non-thermal components at high frequencies (1.4~GHz) decline more slowly than those at low frequencies (141~MHz). This  implies that there are additional mechanisms (i.e. particle acceleration and/or magnetic field amplification via shocks/cold fronts) or inhomogeneous conditions (e.g. stronger turbulence) in the outer regions of the halo as found in the X-ray data. However, our high spectral index uncertainties do not allow us to draw a firm conclusion on this as we find that the spectral index is roughly constant within a radius of 680~kpc.
		
		\item At 141~MHz the relics D and F are connected and are likely to belong to the same source with a projected size of 3.5~Mpc which makes this structure the largest radio relic found to date. Source G is possibly part of this structure which, if included, could increase the relic size to 5.1~Mpc, but deeper observations are needed to confirm this. 
		
		\item The spectral steepening is found across the width of the relics C, D, F, and G, implying the spectral ageing of the sources. Using the spectral indices at the outer edges of the relics, we find that the relics are generated by low-Mach number shocks that are unlikely to directly accelerate thermal electrons to create the observed SB in the relics. An extra-source of energy with steep spectra, e.g. from fossil electrons, is required. The presence of such steep-spectrum sources, i.e. C1 and F1, in our LOFAR data supports this re-acceleration scenario.
		
		\item Thanks to the high-resolution images from the LOFAR 141~MHz and VLA~1.5~GHz, source E is resolved into a double-lobe radio galaxy in the south and an elongated tail in the north, i.e. E1 and E2. We find the spectral index in the region between E1 and E2 flattens, suggesting a possible re-acceleration by a merger shock that is moving outwards in the northern region of E. The spectral steepening in the east-west direction further supports this scenario. 
		
	\end{enumerate}
	
	The late stage merging galaxy cluster ClG0217 is a dynamically rich system that hosts multiple radio relics and a giant radio halo. These extended radio sources are ideal locations for studying particle (re-)acceleration, turbulence and magnetic field amplification in the late stage of cluster merger. Our study using multi-wavelength data sets provide new insights into the nature of the sources (i.e. classification, morphology, spectral properties, and interplay between thermal and non-thermal components). However, due to the quality of the current data, we are unable to confirm some aspects associated with, e.g., radio spectral emission at the X-ray discontinuities in the halo, spectral flattening in the north of source E, and the presence of shock fronts at the location of the relics. Further work using multi-wavelength (radio, X-ray, optical) observations will be necessary to further investigate these.
	
	\begin{acknowledgements}
		
		DNH and AB acknowledge support from the ERC through the grant ERC-Stg DRANOEL n. 714245. 
		MB acknowledges support from the Deutsche Forschungsgemeinschaft under Germany's Excellence Strategy - EXC 2121 "Quantum Universe" - 390833306.
		XZ acknowledges support from China Scholarship Council.
		SRON is supported financially by NWO, The Netherlands Organization for Scientific Research.
		RJvW acknowledges support from the ERC Starting Grant ClusterWeb 804208. 
		AB acknowledges support from the VIDI research programme with project number 639.042.729, which is financed by the Netherlands Organisation for Scientific Research (NWO). 
		AS is supported by the Women In Science Excel (WISE) programme of the Netherlands Organisation for Scientific Research (NWO), and acknowledges the World Premier Research Center Initiative (WPI) and the Kavli IPMU for the continued hospitality. SRON Netherlands Institute for Space Research is supported financially by NWO. 
		GB acknowledges partial support from INAF mainstream program "Galaxy clusters science with LOFAR"
		LOFAR (van Haarlem et al. 2013) is the Low Frequency Array designed and constructed by ASTRON. It has observing, data processing, and data storage facilities in several countries, which are owned by various parties (each with their own funding sources), and that are collectively operated by the ILT foundation under a joint scientific policy. The ILT resources have benefited from the following recent major funding sources: CNRS-INSU, Observatoire de Paris and Université d'Orléans, France; BMBF, MIWF-NRW, MPG, Germany; Science Foundation Ireland (SFI), Department of Business, Enterprise and Innovation (DBEI), Ireland; NWO, The Netherlands; The Science and Technology Facilities Council, UK; Ministry of Science and Higher Education, Poland; The Istituto Nazionale di Astrofisica (INAF), Italy.
		This research made use of the Dutch national e-infrastructure with support of the SURF Cooperative (e-infra 180169) and the LOFAR e-infra group. The J\"ulich LOFAR Long Term Archive and the German LOFAR network are both coordinated and operated by the J\"ulich Supercomputing Centre (JSC), and computing resources on the supercomputer JUWELS at JSC were provided by the Gauss Centre for Supercomputing e.V. (grant CHTB00) through the John von Neumann Institute for Computing (NIC).
		This research made use of the University of Hertfordshire high-performance computing facility and the LOFAR-UK computing facility located at the University of Hertfordshire and supported by STFC [ST/P000096/1], and of the Italian LOFAR IT computing infrastructure supported and operated by INAF, and by the Physics Department of Turin university (under an agreement with Consorzio Interuniversitario per la Fisica Spaziale) at the C3S Supercomputing Centre, Italy.
		The scientific results reported in this article are based  on data obtained from the \textit{Chandra} Data Archive.  This research has made use of software provided by the \textit{Chandra} X-ray Center (CXC) in the application packages CIAO, ChIPS, and Sherpa.
		The National Radio Astronomy Observatory is a facility of the National Science Foundation operated under cooperative agreement by Associated Universities, Inc.
		
	\end{acknowledgements}
	
	\bibliographystyle{aa} % style aa.bst
	\bibliography{library.bib} % your references Yourfile.bib
	
	%\begin{thebibliography}{}
		%
		%  \bibitem[Baker(1966)]{baker} Baker, N. 1966,
		%      in Stellar Evolution,
		%      ed.\ R. F. Stein,\& A. G. W. Cameron
		%      (Plenum, New York) 333
		%
		%\end{thebibliography}

	%
	%%%%%%%%%%%%%%%%%%%%%%%%%%%%%%%%%%%%%%%%%%%%%%%%%%%%%%%%%%%%%%
	%%%%%%%%%%%%%%%%%%%%%%%%%%%%%%%%%%%%%%%%%%%%%%%%%%%%%%%%%%%%%%
	%
	\begin{appendix} %First appendix
		
		%%%%%%%%%%%%%%%%%%%%%%%%%%%%%%%%%%%%
		\section{High-resolution intensity images}
		\label{sec:app_ps}
		
\begin{figure*}[!h]
			\centering
			\includegraphics[width=0.49\textwidth]{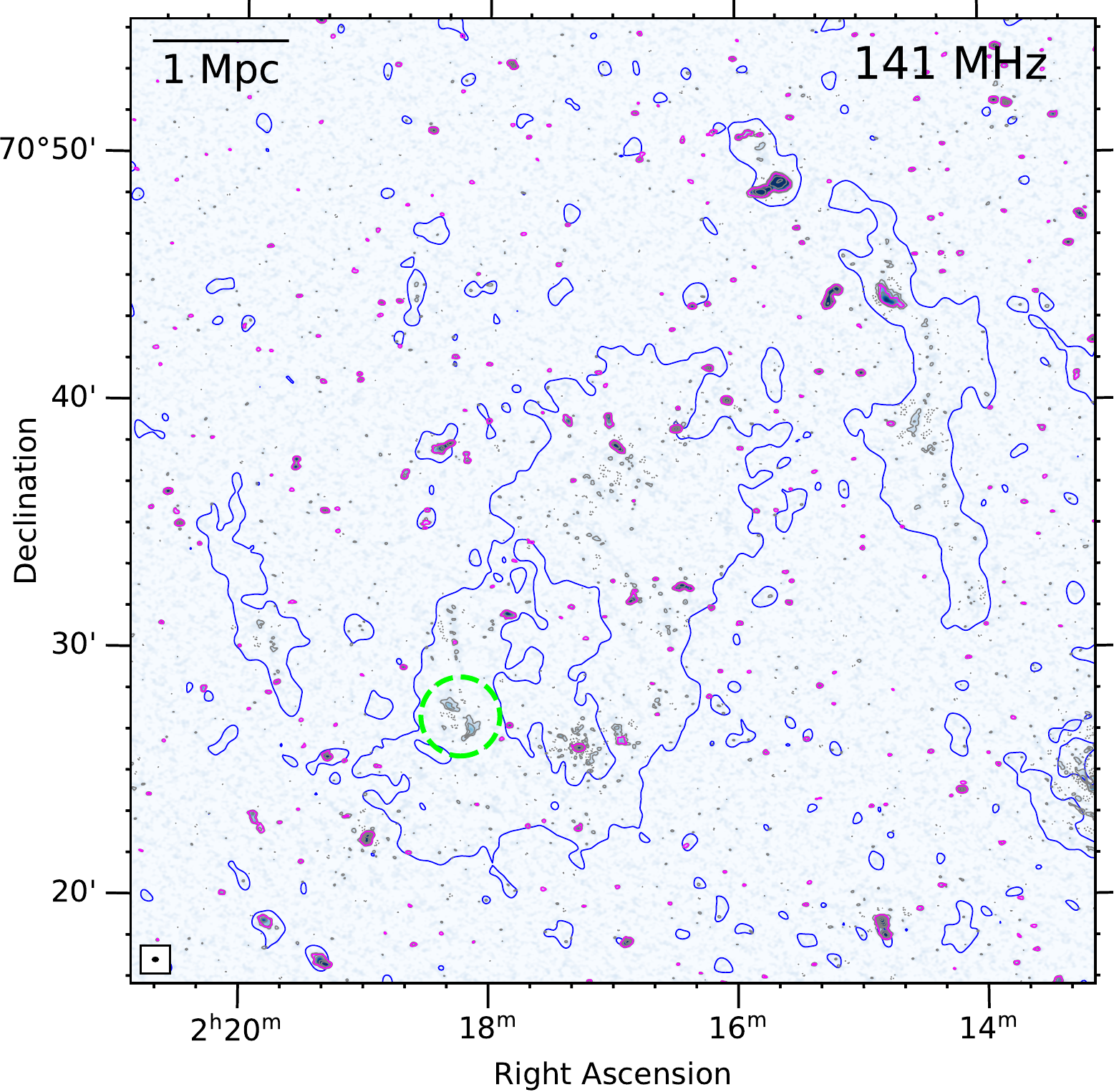} \hfill
			\includegraphics[width=0.49\textwidth]{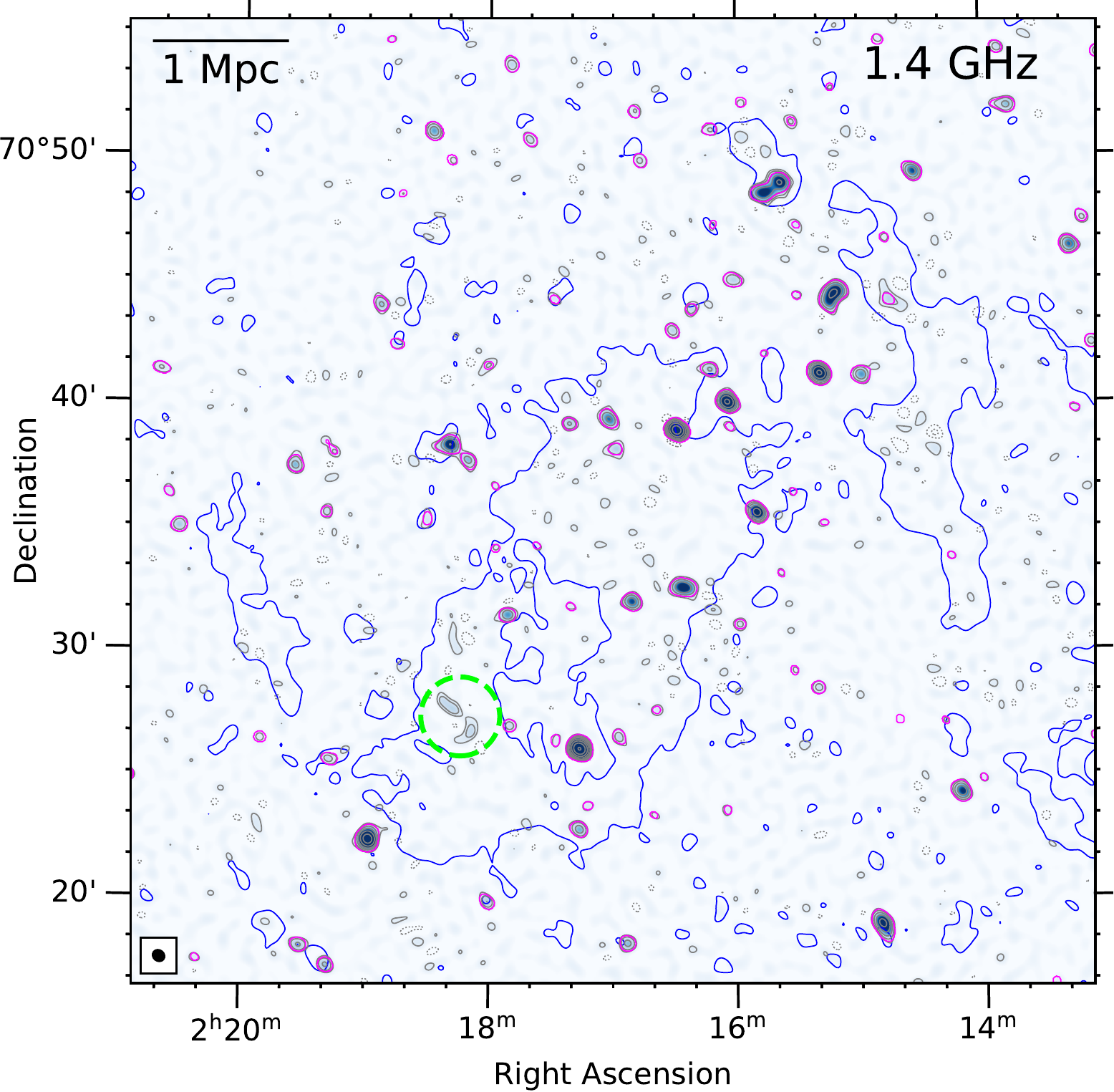} 
			\caption{LOFAR~141~MHz (\textit{left}) and VLA~1.4~GHz (\textit{right}) high-resolution flat-noise images (i.e. primary beam uncorrected). The contaminating discrete sources that are marked with the magenta contours and are removed from the $uv$ data. The source in the dashed circle is not included in the subtracted model. The contours are $[1, 2, 4, 8, 16]\times3\sigma$, where $\sigma_{\rm LOFAR}=140\,\upmu{\rm Jy\,beam}^{-1}$ and $\sigma_{\rm VLA}=80\,\upmu{\rm Jy\,beam}^{-1}$. The compact-source subtracted blue contour is drawn at $2\sigma$, where $\sigma=335\,\upmu{\rm Jy\,beam}^{-1}$ and a beam size of $45.7\arcsec\times44.7\arcsec$.}
			\label{fig:ps}
		\end{figure*}

		The high-resolution intensity images of the cluster ClG0217 are shown in Figure~\ref{fig:ps}. The images are obtained with LOFAR~141~MHz and VLA~1.4GHz data with a combination of $uv$ cut of below $2~k\lambda$ and \texttt{robust} weightings (i.e. $-0.25$ and $-2$ for the LOFAR and VLA data, respectively) to filter out the large-scale emission from the cluster. These images are used to make models of compact sources which are then subtracted from the data. 
		
		%%%%%%%%%%%%%%%%%%%%%%%%%%%%%%%%%%%%%%%
		\section{Fitting the radio halo with a circular 2-dimensional model.}
		\label{sec:app_fitting}
		
\begin{figure*}
			\centering
			\includegraphics[width=0.47\textwidth]{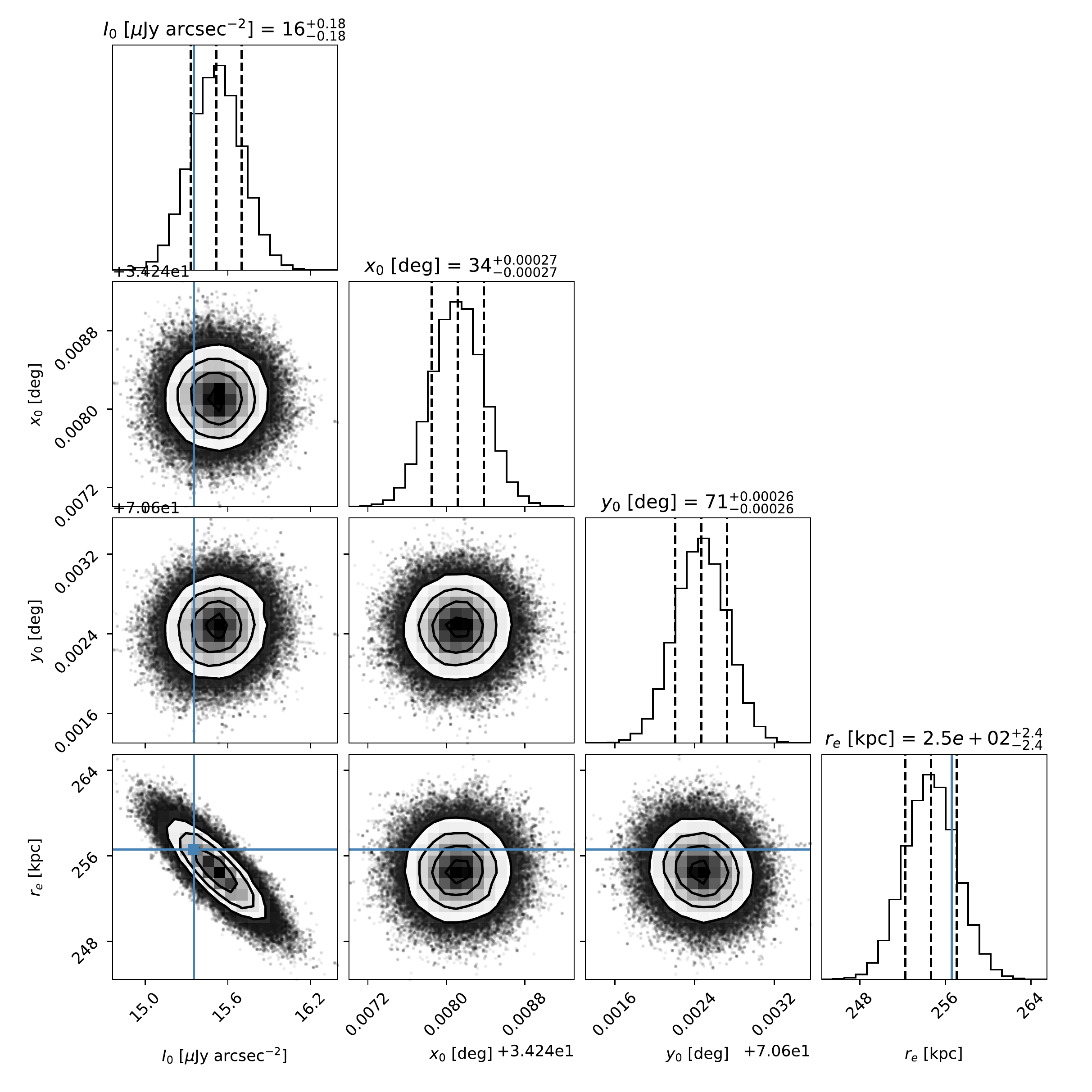} \hfil
			\includegraphics[width=0.47\textwidth]{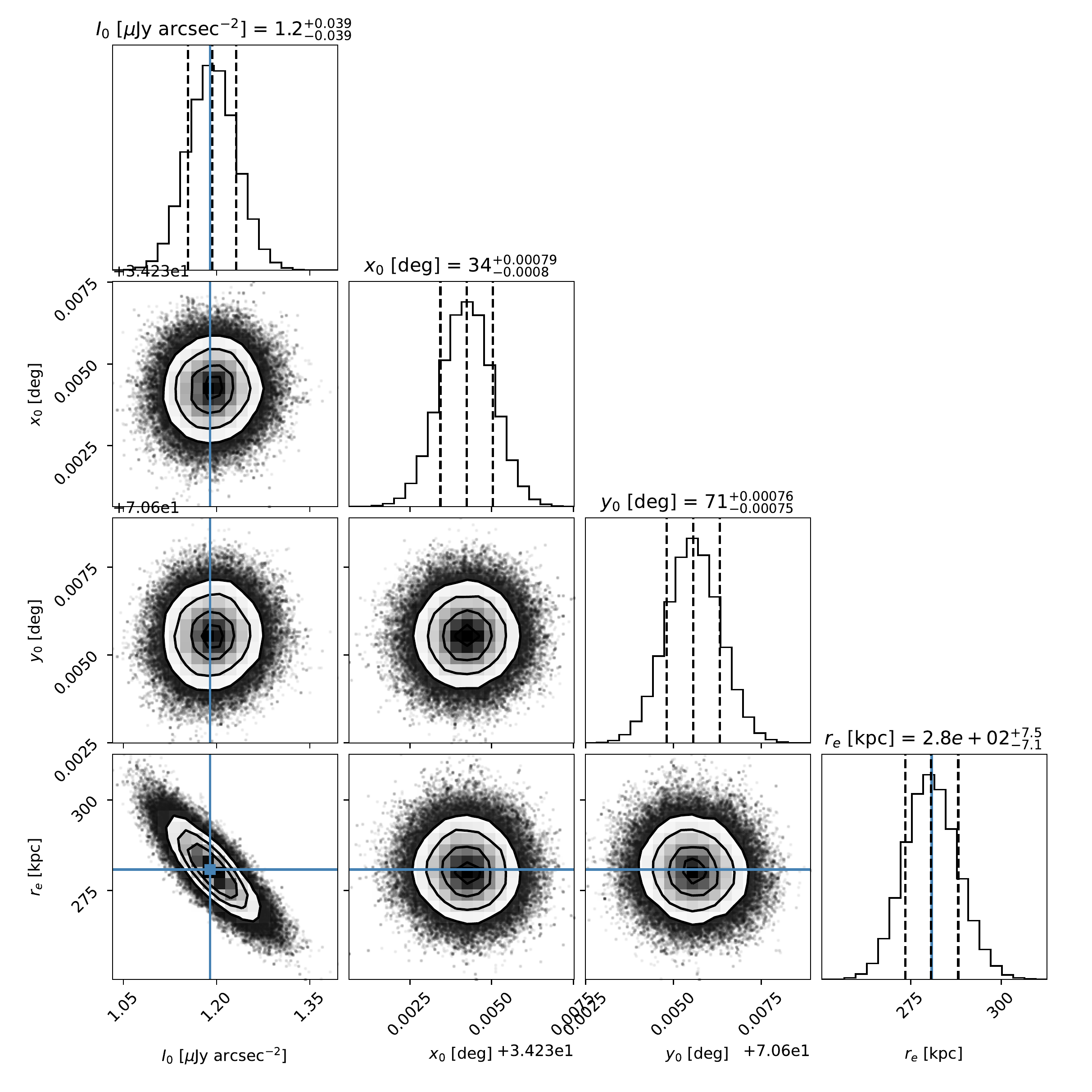} \\
			\caption{Corner plots obtained from the fitting of a 2-dimensional circular model to the SB of the radio halo at 141~MHz (\textit{left}) and 1.4~GHz (\textit{right}). The blue lines show the initial estimates for the peak brightness $I_0$ and the $e$-folding radius $r_e$ by the pre-MCXC fitting. $x_0$ and $y_0$ are the best-fit values in R.A. and Dec. of the halo peak.}
			\label{fig:coner}
		\end{figure*}

		The corner plots for the fitting parameters with circular exponential 2-dimensional model in Section \ref{sec:res_flux} are shown in Figure~\ref{fig:coner}. The maximum likelihood parameters and the uncertainties are estimated using the \texttt{emcee} code \citep{Foreman-Mackey2013}. 
		
		%%%%%%%%%%%%%%%%%%%%%%%%%%%%%%%%%%%%%%%
		\section{Spectral index error maps}
		\label{sec:app_spx_err}
		
\begin{figure*}
			\centering
			\includegraphics[width=0.47\textwidth]{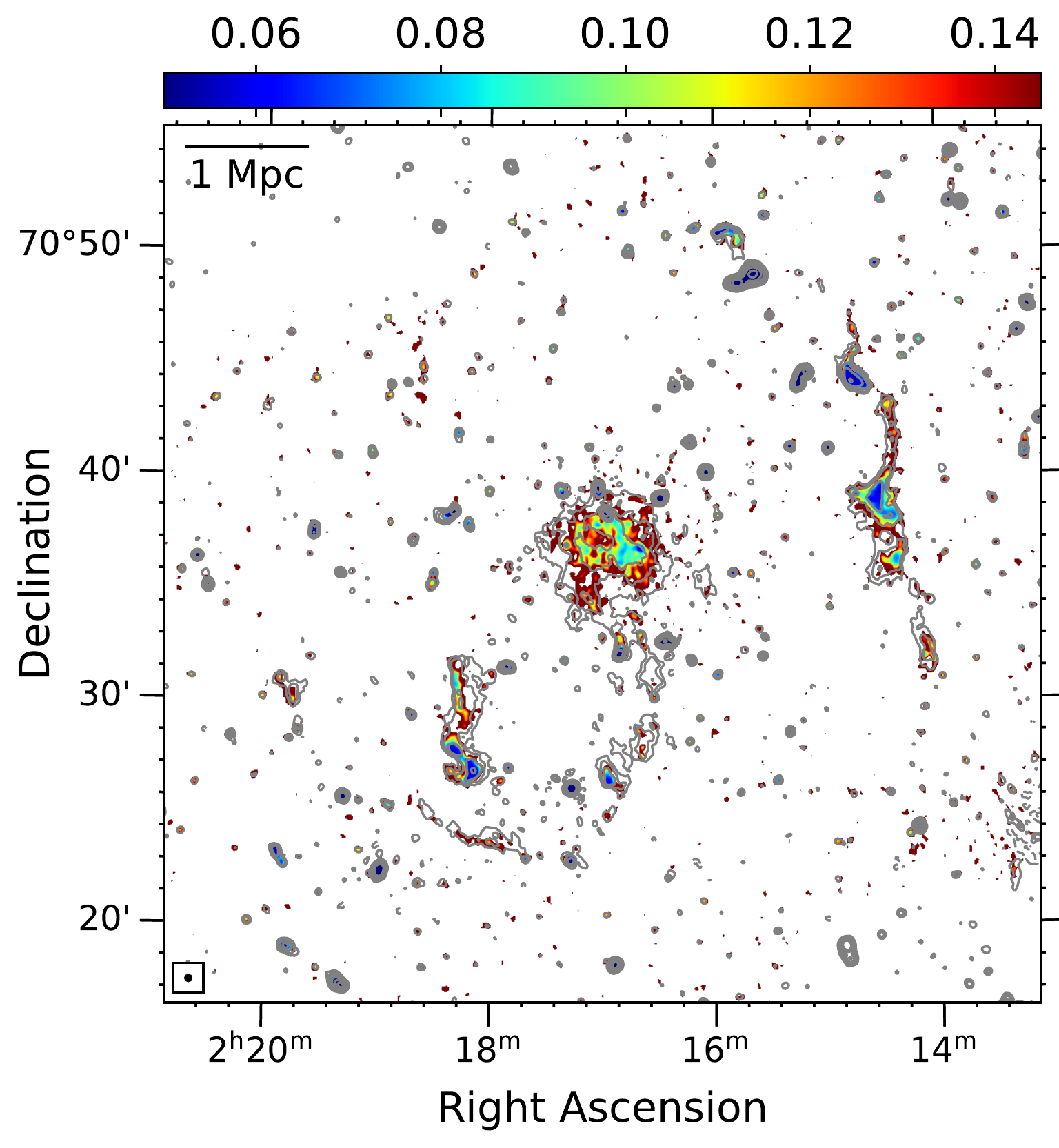} \hfill
			\includegraphics[width=0.47\textwidth]{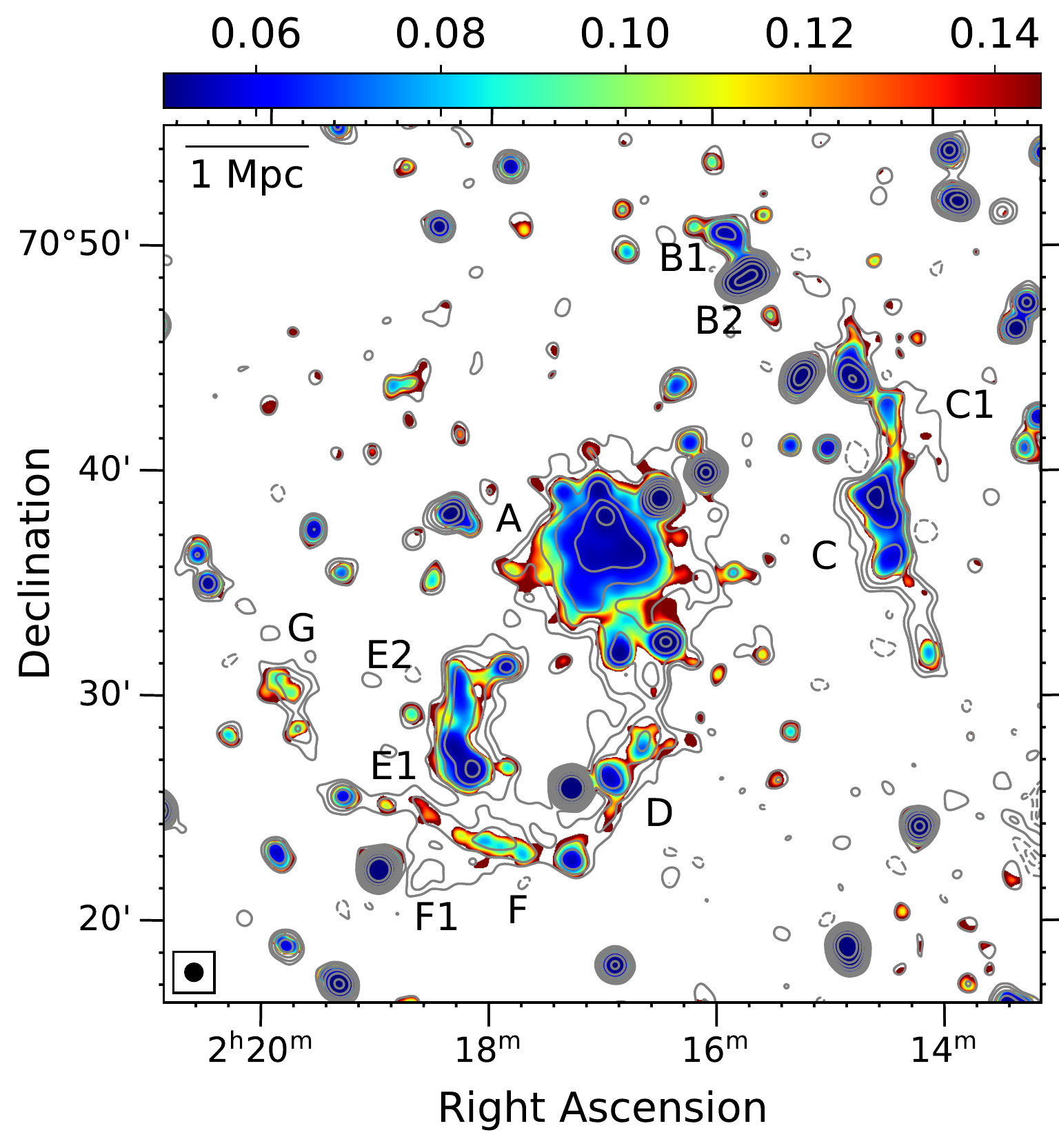} 
			\caption{Spectral index error maps at $16\arcsec$ (\textit{left}) and $46\arcsec$ (\textit{right}) resolutions. The LOFAR 141~MHz contours of $[1, 2, 4, 8, 16]\times3\sigma$, where $\sigma_{\rm LOFAR}=345\,\upmu{\rm Jy\,beam}^{-1}$ (\textit{left}) and $190\,\upmu{\rm Jy\,beam}^{-1}$ (\textit{right}), are overlaid.}
			\label{fig:app_spx_err}
		\end{figure*}

		In Figure~\ref{fig:app_spx_err}, we present the spectral index error maps at the resolution of $16\arcsec$ and $46\arcsec$. The corresponding spectral index maps of ClG0217 are made using the LOFAR and VLA images which are described in Section~\ref{sec:spec}. 
		
		%%%%%%%%%%%%%%%%%%%%%%%%%%%%%%%%%%%%%%%
		\section{Cut-out images}
		\label{sec:app_optical}
		
\begin{figure*}
			\centering
			\includegraphics[width=0.35\textwidth]{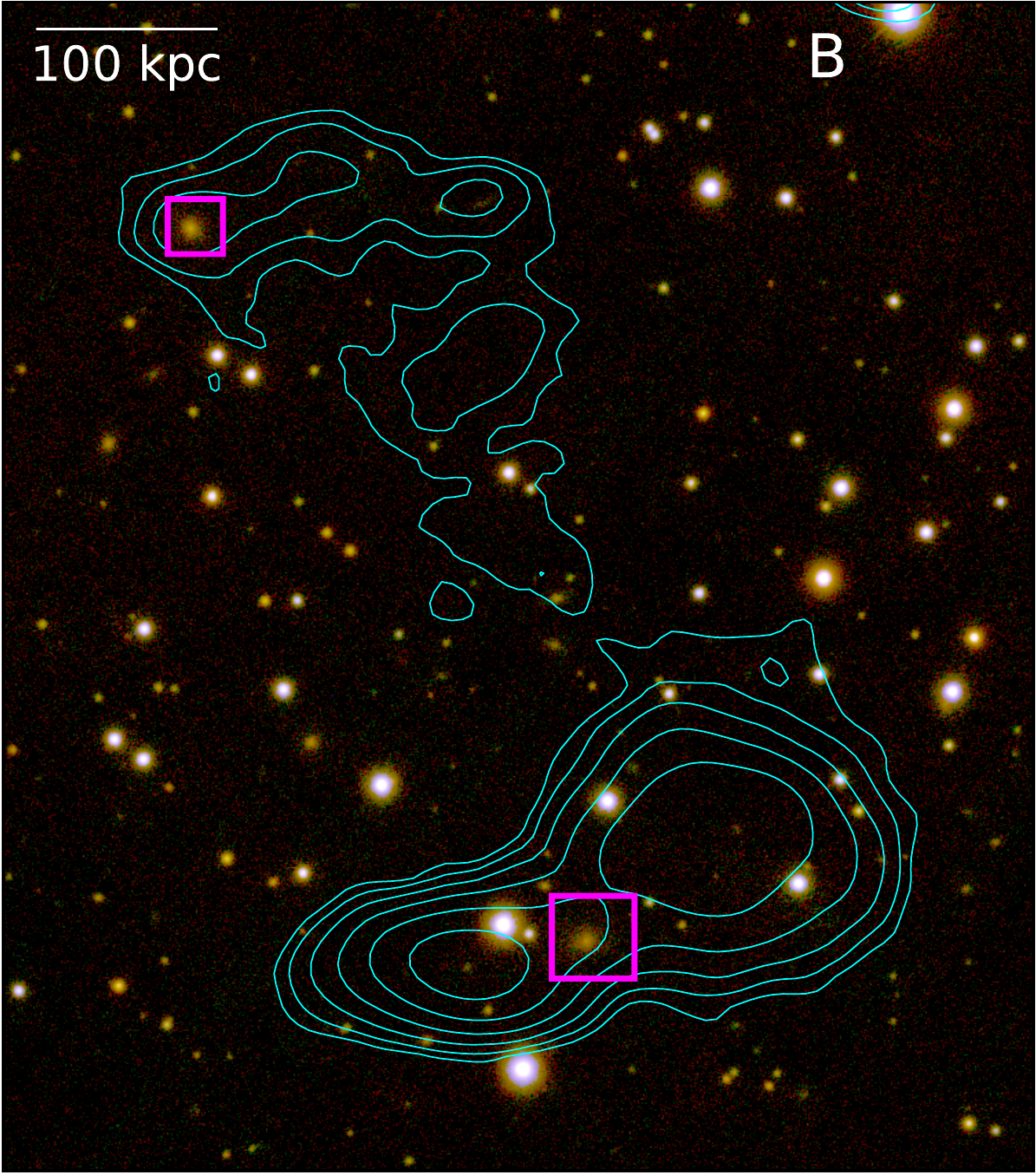} \hfil
			\includegraphics[width=0.35\textwidth]{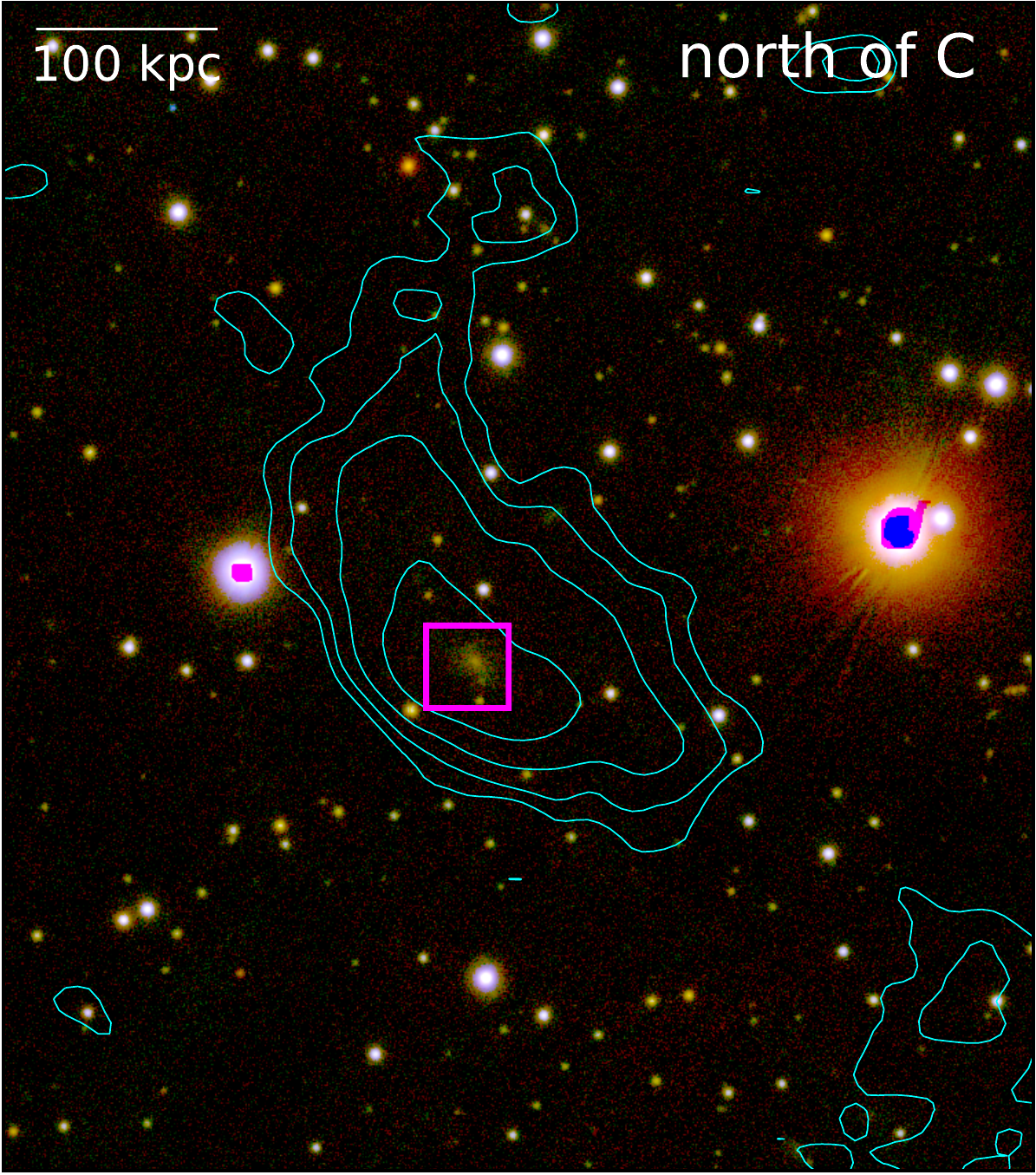} \hfil
			\includegraphics[width=0.27\textwidth]{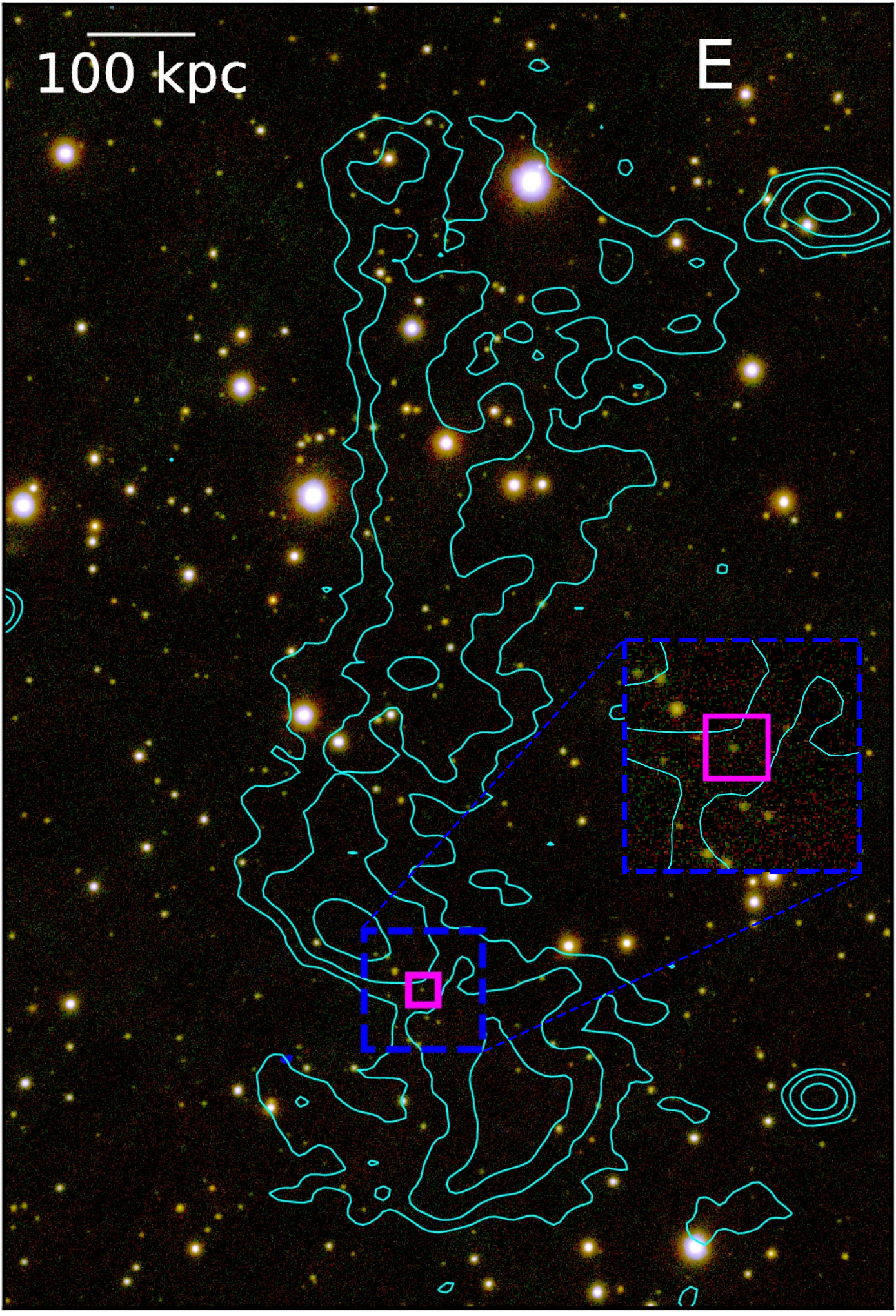}
			\caption{Pan-STARRS colour (y, i, g) images of sources B1+B2 (\textit{left}), C (\textit{middle}), and E (\textit{right}). The square magenta regions show the optical counterparts of the radio galaxies. The overlaid image in the \textit{right} panel shows the central region of the southern part of E (i.e. E1).}
			\label{fig:app_panstarrs}
		\end{figure*}

		To search for optical counterparts of the cluster radio sources, we make cut-out Pan-STARRS colour (band y, i, g) images in the regions of sources B1+B2, C, and E in Figure~\ref{fig:app_panstarrs}. 
		
	\end{appendix}
	
\end{document}